\documentclass[a4paper,fleqn]{cas-dc}

\usepackage[numbers,sort&compress]{natbib}
\usepackage{physics}
\usepackage{siunitx}

\DeclareSIPower\quartic\tothefourth{4}  

\usepackage{amsmath,bm}
\usepackage{framed}
\usepackage{multicol}
\usepackage{nomencl}
\makenomenclature
\usepackage{float}
\usepackage{adjustbox}
\newfloat{nomenclbox}{tp}{lnb}

\renewcommand*\nompreamble{\begin{multicols}{2}}
\renewcommand*\nompostamble{\end{multicols}}

\renewcommand{\nomgroup}[1]{%
  \vspace{1em}%
  \ifstrequal{#1}{A}{\item[\textbf{Latin letters}]}{%
    \ifstrequal{#1}{B}{\item[\textbf{Greek letters}]}{%
      \ifstrequal{#1}{C}{\item[\textbf{Subscripts}]}{%
        \ifstrequal{#1}{D}{\item[\textbf{Acronyms}]}{}%
      }%
    }%
  }%
}
\usepackage[version=4]{mhchem}
\usepackage{caption}
\usepackage{parskip}
\usepackage{subfig}
\usepackage{enumerate}

\usepackage{threeparttable}
\usepackage{mathtools}
\usepackage{booktabs}

\expandafter\def\csname __first_footerline:\endcsname{%
  \begingroup\small\sffamily
  \ifnum\value{blind}>0\relax\else
    \csname __short_authors:\endcsname:~%
  \fi
  {\rmfamily\itshape Preprint}%
  \endgroup}

\usepackage{xcolor}

\definecolor{myred}{RGB}{238,52,36}
\definecolor{mypurple}{RGB}{146,45,145}
\definecolor{myskin}{RGB}{238,202,178}

\usepackage[draft]{changes}
\definechangesauthor[name={J. Valdivia}, color={blue}]{JV}
\definechangesauthor[name={P. Pinto}   , color={orange}]{PP}
\definechangesauthor[name={J. Urban}   , color={red}]{JU}
\begin{document}

\let\WriteBookmarks\relax
\def\floatpagepagefraction{1}
\def\textpagefraction{.001}
\shorttitle{2D drying and oxidative pyrolysis of wood cylinders}
\shortauthors{A. Bhattacharjee et~al.}

\title [mode = title]{Two-dimensional modeling of coupled drying and oxidative pyrolysis of wood cylinders under convective-radiative heating}

\author[]{Anurag Bhattacharjee}[type=editor,
                        prefix=,
                        role=,
                        orcid=0009-0004-4496-0414]
\ead{abhattacharjee@wpi.edu}
\cormark[1]
\affiliation[]{organization={Department of Fire Protection Engineering, Worcester Polytechnic Institute},
                addressline={100 Institute Rd}, 
                city={Worcester},
                state={MA 01609},
                country={USA}}

\author[]{Jorge Valdivia}[orcid=0009-0003-4251-1108]

\author[]{James L. Urban}[orcid=0000-0002-2476-8212]
\ead{jurban@wpi.edu}
\cormark[1]

\cortext[cor1]{Corresponding author}

\begin{abstract}
The moisture content and heating environment of woody fuels govern when they become available to burn and how they ignite. Dead fuel moisture models track drying under weather but not the decomposition that follows, while pyrolysis models represent drying with simple evaporation laws developed for fast heating. This study presents a two-dimensional polar ($r$-$\theta$) heat and mass transfer model of a wood cylinder that treats both in one framework. Moisture moves as bound water through the cell walls and as vapor through the pores, and the two exchange at a finite rate driven by a sorption isotherm, so the same formulation describes slow drying in air and rapid drying under fire exposure. The moisture model is coupled to thermal and oxidative pyrolysis and char oxidation, and the surface boundary conditions resolve the angular variation of convective heating around a cylinder in cross flow and of one-sided radiative heating. The model is first applied to the drying of \SIlist{5;15;50;150;500}{\milli\meter} diameter cylinders from an initial moisture content of 10\% (dry basis) in hot (\SI{37.8}{\celsius}), dry (0\% relative humidity) air at \SI{22.4}{\meter\per\second}. Drying is controlled by internal moisture diffusion: the drying time scales with the square of the diameter, the predicted timelags fall within the diameter ranges of the standard fuel timelag classes, and halving the external mass transfer coefficient lengthens drying by less than 8\%. The model is then applied to oven-dried white oak dowels of \SI{6.35}{\milli\meter} diameter exposed to radiant fluxes up to \SI{42}{\kilo\watt\per\meter\squared} and convective cross flows at gas temperatures up to \SI{430}{\celsius}. Predicted glowing ignition times lie within one standard deviation of the measurements, and the measured mean flaming ignition times fall within the predicted ignition windows. Under coupled heating, radiation produces steep surface temperature gradients and one-sided char and ash fronts. An initial moisture content of 20\% delays glowing ignition by about 30\%.
\end{abstract}


\maketitle

\begin{nomenclbox*}[!t]
\begin{adjustbox}{max totalheight=0.85\textheight, max width=\textwidth}
\fbox{\begin{minipage}{0.97\textwidth}
\vspace{4pt}

\nomenclature[A]{$A$}{Area (\si{\meter\squared})}
\nomenclature[A]{$\dot{D}_{\mathrm{O_2}}$}{Oxygen demand rate (kg$\cdot$m$^{-3}\cdot$s$^{-1}$)}
\nomenclature[A]{$\dot{S}_{\mathrm{O_2}}$}{Oxygen supply rate (kg$\cdot$m$^{-3}\cdot$s$^{-1}$)}
\nomenclature[B]{$\lambda$}{Oxygen limiter factor ($-$)}
\nomenclature[A]{$F$}{View factor ($-$)}
\nomenclature[A]{$G$}{Irradiation (\si{\watt\per\meter\squared})}
\nomenclature[A]{$L$}{Rod length (\si{\meter})}
\nomenclature[A]{$\mathrm{Sc}$}{Schmidt number ($-$)}
\nomenclature[A]{$U$}{Free-stream velocity (\si{\meter\per\second})}
\nomenclature[A]{$V$}{Volume of a control volume (\si{\meter\cubed})}
\nomenclature[A]{$c_p$}{Specific heat capacity (\si{\joule\per\kilogram\per\kelvin})}
\nomenclature[A]{$D$}{Cylinder diameter (\si{\meter})}
\nomenclature[A]{$\mathcal{D}_X$}{Bound-water diffusivity (\si{\meter\squared\per\second})}
\nomenclature[A]{$\mathcal{D}$}{Gas-phase diffusivity (\si{\meter\squared\per\second})}
\nomenclature[A]{$E$}{Activation energy (\si{\joule\per\mole})}
\nomenclature[A]{$h$}{Specific enthalpy (\si{\joule\per\kilogram}); heat transfer coefficient (\si{\watt\per\meter\squared\per\kelvin})}
\nomenclature[A]{$j_X$}{Moisture flux (\si{\meter\per\second})}
\nomenclature[A]{$h_d$}{Mass transfer coefficient (\si{\meter\per\second})}
\nomenclature[A]{$\Delta h_v$}{Latent heat of vaporization (\si{\joule\per\kilogram})}
\nomenclature[A]{$\Delta H$}{Heat of reaction (\si{\joule\per\kilogram})}
\nomenclature[A]{$k$}{Thermal conductivity (\si{\watt\per\meter\per\kelvin})}
\nomenclature[A]{$K$}{Permeability (\si{\meter\squared})}
\nomenclature[A]{$\dot m''$}{Mass flux (\si{\kilogram\per\meter\squared\per\second})}
\nomenclature[A]{$M$}{Molar mass (\si{\kilogram\per\mole})}
\nomenclature[A]{$n$}{Reaction order ($-$)}
\nomenclature[A]{$\mathrm{Nu}$}{Nusselt number ($-$)}
\nomenclature[A]{$P$}{Pressure (\si{\pascal})}
\nomenclature[A]{$\dot q''$}{Heat flux (\si{\watt\per\meter\squared})}
\nomenclature[A]{$\dot Q'''$}{Volumetric heat release rate (\si{\watt\per\meter\cubed})}
\nomenclature[A]{$r$}{Radial coordinate (\si{\meter})}
\nomenclature[A]{$R$}{Cylinder radius (\si{\meter})}
\nomenclature[A]{$\mathcal{R}$}{Universal gas constant (\si{\joule\per\mole\per\kelvin})}
\nomenclature[A]{$\mathrm{Re}$}{Reynolds number ($-$)}
\nomenclature[A]{$\mathrm{Sh}$}{Sherwood number ($-$)}
\nomenclature[A]{$t$}{Time (\si{\second})}
\nomenclature[A]{$T$}{Temperature (\si{\kelvin})}
\nomenclature[A]{$T_f$}{Film temperature (\si{\kelvin})}
\nomenclature[A]{$X$}{Moisture content, dry basis ($-$)}
\nomenclature[A]{$Y$}{Mass fraction ($-$)}
\nomenclature[A]{$Z$}{Pre-exponential factor (\si{\per\second})}
\nomenclature[A]{$A'''$}{Interfacial area per unit volume (\si{\per\meter})}
\nomenclature[A]{$C_b$}{Packed-bed shape factor ($-$)}
\nomenclature[A]{$d_f$}{Fiber diameter (\si{\meter})}
\nomenclature[A]{$H, H_1, H_2$}{Hailwood-Horrobin constants ($-$)}
\nomenclature[A]{$\mathrm{Re}_p$}{Pore Reynolds number ($-$)}
\nomenclature[A]{$\mathrm{Sh}_p$}{Pore Sherwood number ($-$)}
\nomenclature[A]{$W$}{Hailwood-Horrobin molecular weight per sorption site (\si{\gram\per\mole})}
\nomenclature[B]{$\xi_w$}{Fraction of the transport path in cell walls ($-$)}
\nomenclature[B]{$\delta$}{Target change per time step}
\nomenclature[C]{$p$}{Pore}
\nomenclature[A]{$a_X$}{Moisture diffusivity anisotropy ratio, $\mathcal{D}_{X,\theta}/\mathcal{D}_{X,r}$ ($-$)}

\nomenclature[B]{$\beta$}{Evaporation exchange coefficient (\si{\per\second})}
\nomenclature[B]{$\chi$}{Volume fraction ($-$)}
\nomenclature[B]{$\varepsilon$}{Emissivity ($-$)}
\nomenclature[B]{$\mu$}{Dynamic viscosity (\si{\pascal\second})}
\nomenclature[B]{$\tau$}{Tortuosity ($-$)}
\nomenclature[B]{$\vartheta$}{Time-weighting factor ($-$)}
\nomenclature[B]{$\theta$}{Angular coordinate (\si{\radian})}
\nomenclature[B]{$\nu$}{Kinematic viscosity (\si{\meter\squared\per\second})}
\nomenclature[B]{$\rho$}{Bulk density (\si{\kilogram\per\meter\cubed})}
\nomenclature[B]{$\sigma$}{Stefan-Boltzmann constant (\si{\watt\per\meter\squared\per\kelvin\tothe{4}})}
\nomenclature[B]{$\phi$}{Relative humidity / water activity ($-$)}
\nomenclature[B]{$\psi$}{Porosity ($-$)}
\nomenclature[B]{$\dot\omega'''$}{Volumetric reaction rate (\si{\kilogram\per\meter\cubed\per\second})}

\nomenclature[C]{$d$}{Destruction}
\nomenclature[C]{$e$}{Evaporation}
\nomenclature[C]{$f$}{Formation}
\nomenclature[C]{$g$}{Gas phase}
\nomenclature[C]{$i$}{Condensed-phase species}
\nomenclature[C]{$j$}{Gas-phase species}
\nomenclature[C]{$k$}{Heterogeneous reaction}
\nomenclature[C]{$\ell$}{Homogeneous reaction}
\nomenclature[C]{$s$}{Surface}
\nomenclature[C]{$\mathrm{sat}$}{Saturation}
\nomenclature[C]{$b$}{Bound water; boiling point}
\nomenclature[C]{$d$}{Dry solid (on $\overline{\rho}$, $\overline{k}$, $\overline{c}_p$)}

\nomenclature[C]{$\mathrm{fsp}$}{fiber saturation point}
\nomenclature[C]{$w$}{Water}
\nomenclature[C]{$v$}{Virgin (dry) wood}
\nomenclature[C]{$\infty$}{Ambient}

\nomenclature[D]{FMC}{Fuel moisture content}
\nomenclature[D]{GCI}{Grid convergence index}
\nomenclature[D]{RH}{Relative humidity}

{\footnotesize
\setlength{\parskip}{0pt}
\printnomenclature

}

\vspace{4pt}
\end{minipage}}
\end{adjustbox}
\end{nomenclbox*}

%%%%%%%%%%%%%%%%%%%%%%%%%%%%%%%%%%%%%%%%%%%%%%%%%%%%%%%%%%%%%%%%%%%%%
\section{Introduction}
\label{sec:intro}

The moisture content of dead woody fuels is one of the strongest controls on whether a wildland or wildland-urban interface fire ignites, how fast it spreads, and how much fuel it consumes~\cite{Anderson1982}. Depending on their configuration, dead fuels exchange moisture with the air or soil.  In both cases their moisture content is strongly influenced by the weather, but with a delay that increases with dead fuel element size. Fire danger rating systems capture this through timelag classes: woody fuels below \SI{6.4}{\milli\meter} in diameter are classified as 1-h fuels, \SIrange{6.4}{25}{\milli\meter} as 10-h fuels, \SIrange{25}{76}{\milli\meter} as 100-h fuels, and larger material as 1000-h fuel~\cite{Anderson1982}, where the timelag is loosely the time a particle needs to move about two-thirds of the way to equilibrium with its environment~\cite{bradshaw1984nfdrs}. Under hot, dry and windy fire weather, fine fuels can lose most of their moisture within hours, while large fuels remain comparatively wet for days to weeks. This difference sets which fuels are available to burn at the onset of an event.

Process-based dead fuel moisture models resolve heat and moisture transport inside an idealized fuel stick. The model of Nelson~\cite{nelson2000prediction} solves coupled heat and moisture equations within a cylindrical 10-h stick driven by air temperature, humidity, solar radiation and rainfall, and can be adapted to sticks of other sizes. Matthews~\cite{matthews2014dead} reviews the wider body of empirical and process-based models. These models are one-dimensional in radius and are built to track moisture over diurnal and seasonal cycles. They are not coupled to the thermal decomposition, oxidation and ignition that follow once the same fuel is exposed to a fire.

Fire-science pyrolysis models address that second stage. Thermal degradation models such as Gpyro~\cite{Lautenberger2009,Lautenberger2009_FSJ}, later extended to three dimensions~\cite{lautenberger2014gpyro3d}, solve condensed-phase species, gas transport through the pores, pressure and energy with multi-step kinetics, and reviews of wood and biomass pyrolysis modeling summarize the many kinetic and transport sub-models in use~\cite{di2008modeling}. Drying in these models is commonly represented either as an equilibrium process pinned at the boiling point or as an Arrhenius-type evaporation reaction, both developed for heating rates far higher than ambient weathering. 

A cylindrical fuel element in a crossflow is not uniformly heated~\cite{Incropera2018,Zukauskas1987}. The local convective heat and mass transfer coefficients vary strongly around the circumference between the forward stagnation point and the wake. Similarly, often radiation heating is directional only exposing one side to radiation.
One-dimensional radial models average these effects away, yet surface temperature, local drying and the location of first ignition all can depend on the local boundary conditions. In addition, moisture and pyrolysis models have largely been developed in separate communities, with drying under weather and drying under fire exposure treated as different problems, although they are the same physics acting on different time scales.

This work presents a two-dimensional ($r$-$\theta$) model of a moist wood cylinder that spans both regimes in one framework. Bound water diffuses through the cell walls with a moisture and temperature-dependent diffusivity~\cite{siau2012transport,gezici2017bound} and exchanges with the pore gas at a finite rate, driven by the difference between the equilibrium vapor density given by a Hailwood-Horrobin sorption isotherm~\cite{hailwood1946absorption,ross2010wood} and the vapor density in the pores. The same exchange law applies inside the solid, where it maintains local sorption equilibrium, and at the surface, where the external boundary layer limits it. Water vapor, oxygen and pyrolysis gases are transported through the pores by Darcy flow and diffusion. The energy equation includes conduction, latent heat and reaction heat, with an angularly resolved surface boundary condition that combines local heat and mass transfer distributions for a cylinder in cross flow~\cite{Zukauskas1987}, a blowing correction, and radiation from a heater and heated surroundings. Thermal and oxidative pyrolysis, char oxidation and the oxidation of pyrolysis gases in the pores follow multi-step Arrhenius kinetics. The equations are discretized by finite differences on a polar grid and advanced implicitly with an adaptive time step.

The model is applied to two problems at opposite ends of the time-scale range. First, the drying of \SIlist{5;15;50;150;500}{\milli\meter} diameter wood cylinders is simulated, spanning the 1-h, 10-h, 100-h and largest timelag classes, from an initial fuel moisture content of 10\% (dry basis) in hot (\SI{37.8}{\celsius}), completely dry (0\% relative humidity) air moving at \SI{22.4}{\meter\per\second}. A mesh convergence study on the \SI{50}{\milli\meter} case sets the grid resolution. Second, we simulate oven-dried white oak dowels of \SI{6.35}{\milli\meter} diameter under convective heating and under coupled convective and radiative heating, and compare predicted glowing ignition times with experiments~\cite{Valdivia_thesis}.

The contributions of this paper are: (i) a two-dimensional formulation coupling bound-water and pore-vapor transport through evaporation-condensation exchange with oxidative pyrolysis and char oxidation in a cylinder under asymmetric convective-radiative heating, with angularly resolved surface heat and mass transfer; (ii) a numerical framework and verification procedure for this stiff, multi-time-scale system on a polar grid; (iii) predictions of how drying time scales with fuel diameter and wind speed under extreme fire weather; and (iv) a comparison of predicted and measured glowing and flaming ignition times for small-diameter dowels. Section~\ref{sec:model} presents the governing equations, Section~\ref{sec:numerics} the numerical framework, Section~\ref{sec:drying} the drying study, Section~\ref{sec:ignition} the ignition study, Section~\ref{sec:discussion} the limitations of the model, and Section~\ref{sec:conclusions} the conclusions.

%%%%%%%%%%%%%%%%%%%%%%%%%%%%%%%%%%%%%%%%%%%%%%%%%%%%%%%%%%%%%%%%%%%%%
\section{Model formulation}
\label{sec:model}

\subsection{Problem description and assumptions}

\begin{figure}
    \centering
    \includegraphics[width=\linewidth]{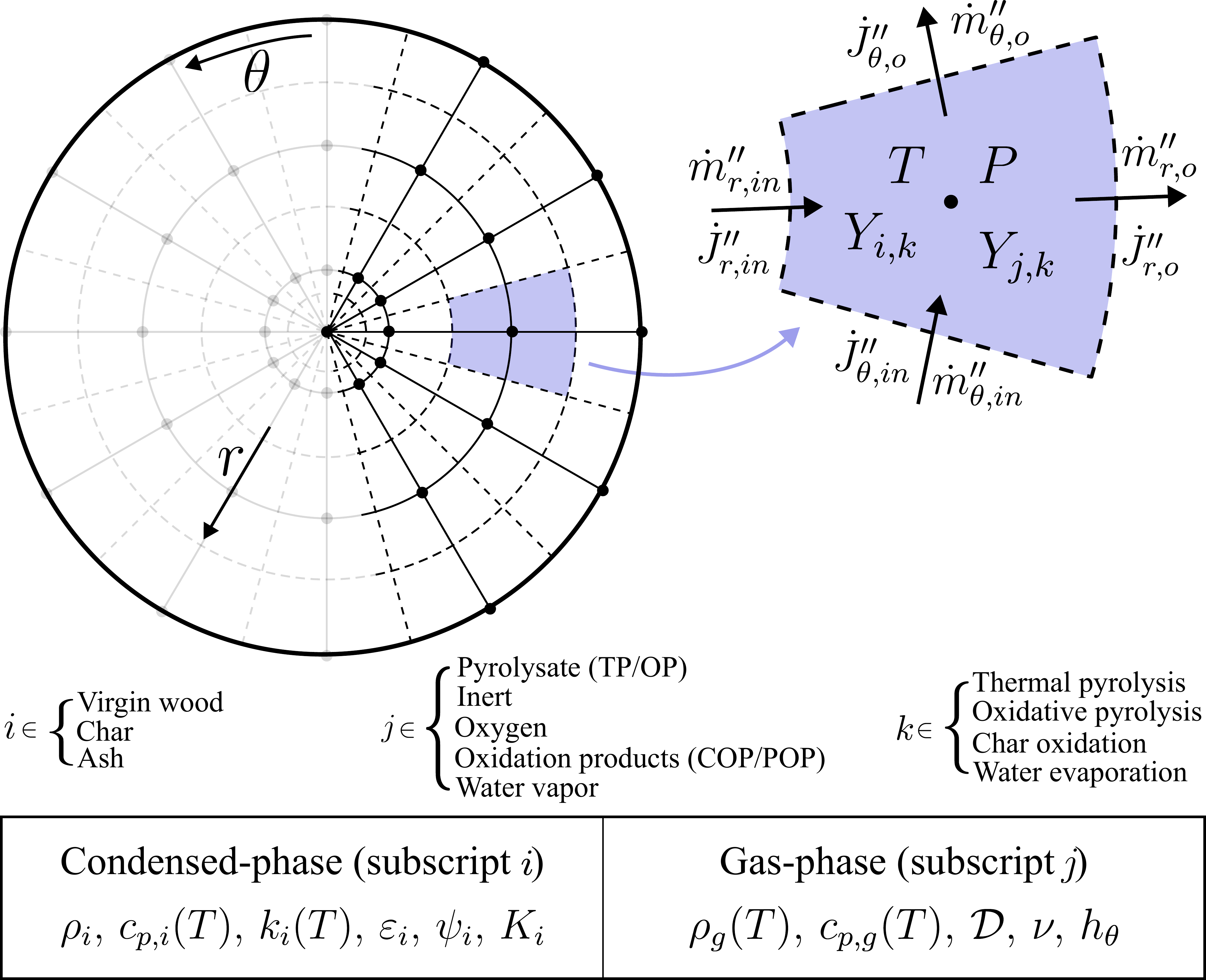}
    \caption{Cross-section of the wood cylinder showing the polar coordinate system.}
    \label{fig:schematic}
\end{figure}

A porous media heat and mass transfer model with the capability to resolve pyrolysis reactions and vapor-liquid phase change was developed following the Generalized Pyrolysis (Gpyro) framework~\cite{Lautenberger2009,Lautenberger2009_FSJ,lautenberger2014gpyro3d} and applied to a two-dimensional polar (radial-angular) geometry to resolve circumferential variations in external heating over the cylinder surface (Fig.~\ref{fig:schematic}). The angle $\theta$ is measured from the forward stagnation point of the cross flow. Because the thermal heating is uniform along the cylinder axis, axial gradients are neglected, which reduces the problem to the $r$-$\theta$ plane. The model is extended relative to Ref.~\cite{Lautenberger2009} by a transport equation for bound water and a finite-rate evaporation-condensation exchange between the walls of the pores and the pore gas (Section~\ref{sec:moisture}).

The following modeling assumptions were adopted:
\begin{enumerate}
    \item Heating and transport are uniform in the axial direction (infinite cylinder).
    \item Condensed-phase species have temperature-dependent heat capacity ($c_{p,i}$) and conductivity ($k_i$). The bulk density ($\rho_i$), emissivity ($\varepsilon_i$), permeability ($K_i$), porosity ($\psi_i$) and tortuosity ($\tau_i$) of each species are constant, so the mixture values change only with composition.
    \item Overbarred quantities (e.g., $\overline{\rho}_d$) denote mass or volume- weighted averages.
    \item Gas-phase properties ($\rho_g$, $\mu_g$, $c_{p,g}$, $k_g$) are temperature dependent, and the mean molar mass of the pore gas is approximated by that of air.
    \item Bulk gas transport through the porous medium follows Darcy's law.
    \item The gas and solid phases are in local thermal equilibrium ($T_g = T$).
    \item The solid matrix does not shrink or swell (isochoric).
    \item Water is held in the cell walls as bound water and moves by diffusion, it leaves the solid only as vapor. Liquid (free-water) flow above the fiber saturation point is not modeled.
\end{enumerate}

Subscripts denote phase and reaction indices. The index $i$ represents condensed-phase species ($i \in \{\text{virgin wood } (v), \\ \text{char } (c), \text{ash } (a)\}$) and $j$ denotes the seven gas-phase species: thermal pyrolysate, \ce{N2} (inert), \ce{O2}, oxidative pyrolysate, char-oxidation products, pyrolysate-oxidation products and water vapor. The heterogeneous processes $k$ are:
\begin{enumerate}
    \item Dry wood $\to$ char + gas (\textit{thermal pyrolysis}),
    \item Dry wood + \ce{O2} $\to$ char + gas (\textit{oxidative pyrolysis}),
    \item Char + \ce{O2} $\to$ ash + gas (char oxidation),
    \item Bound water $\rightleftharpoons$ water vapor (\textit{evaporation and condensation}).
\end{enumerate}
Evaporation (Process 4) is treated like a reaction that converts bound water into water vapor. Its rate, $\dot\omega'''_e$, is written separately from the pyrolysis reactions (Section~\ref{sec:moisture}). It removes water from the moisture equation, adds vapor to the gas mass, vapor and pressure equations, and absorbs latent heat in the energy equation. A negative $\dot\omega'''_e$ represents condensation.

\subsection{Composition and properties of the moist solid}
\label{sec:props}

The dry solid in each control volume is described by the volume fractions $\chi_i$ of virgin wood, char and ash ($\sum_i \chi_i = 1$), from which the dry bulk density and mass fractions are
\begin{equation}
    \overline{\rho}_d = \sum_i \chi_i\,\rho_i, \qquad Y_i = \frac{\chi_i\,\rho_i}{\overline{\rho}_d}.
\end{equation}
Bound water is described by the dry-basis moisture content $X$, defined relative to the bulk density of virgin wood, so that the water mass per unit volume is $X\rho_v$. The bulk density of the moist solid and the mass fraction of water in it are
\begin{equation}
    \overline{\rho} = \overline{\rho}_d + X\rho_v, \qquad Y_w = \frac{X\rho_v}{\overline{\rho}}.
\end{equation}
Dry-solid properties are averaged as
\begin{gather}
    \overline{k}_d = \sum_i \chi_i k_i, \qquad \overline{c}_{p,d} = \sum_i Y_i c_{p,i}, \nonumber\\
    \overline{\varepsilon}_d = \sum_i \chi_i \varepsilon_i, \qquad \overline{\psi} = \sum_i \chi_i \psi_i, \\
    \overline{K} = \sum_i \chi_i K_i, \qquad \overline{\tau} = \sum_i \chi_i \tau_i, \nonumber
\end{gather}
Water content is accounted through mass weighting,
\begin{align}
    \overline{c}_p &= (1-Y_w)\,\overline{c}_{p,d} + Y_w\,c_{p,w}(T), \nonumber\\
    \overline{k} &= (1-Y_w)\,\overline{k}_d + Y_w\,k_w(T), \\
    \overline{\varepsilon} &= (1-Y_w)\,\overline{\varepsilon}_d + Y_w\,\varepsilon_w, \nonumber
\end{align}
where $c_{p,w}(T)$ and $k_w(T)$ are polynomial fits for liquid water (Table~\ref{tab:water}) and $\varepsilon_w$ is the emissivity of water. Because bound water resides in the cell walls, the porosity $\overline{\psi}$ available to the pore gas is not altered by moisture. 

The conductivity and specific heat of each condensed-phase species vary with temperature as~\cite{Lautenberger2009}
\begin{gather}
    k_i(T) = k_{0,i}\left(\frac{T}{T_r}\right)^{n_{k,i}} + \gamma_i\,\sigma T^3, \nonumber\\
    c_{p,i}(T) = c_{0,i}\left(\frac{T}{T_r}\right)^{n_{c,i}},
    \label{eq:prop_T}
\end{gather}
where $T_r = \SI{295.65}{\kelvin}$ and the term $\gamma_i\sigma T^3$ accounts for radiation across the pores. The property values are listed in Table~\ref{tab:cond_props}.

\begin{table*}[!t]
\centering
\caption{Condensed-phase properties. The bulk density of virgin wood is $\rho_v = (1-\psi_v)\rho_{s}$ with the cell-wall density $\rho_s = \SI{1317.6}{\kilogram\per\meter\cubed}$. The tortuosity is $\tau_i = 1 + 0.01\rho_i$. Bound water has emissivity $\varepsilon_w = 0.77$.}
\label{tab:cond_props}
\small
\setlength{\tabcolsep}{4pt}
\begin{tabular}{@{}lccccccccc@{}}
\toprule
Species $i$ & $k_0$ (\si{\watt\per\meter\per\kelvin}) & $n_k$ & $\rho$ (\si{\kilogram\per\meter\cubed}) & $c_0$ (\si{\joule\per\kilogram\per\kelvin}) & $n_c$ & $\varepsilon$ & $\gamma$ (\si{\meter}) & $K$ (\si{\meter\squared}) & $\psi$ \\
\midrule
Virgin wood ($v$) & 0.176 & 0.594 & 790.6 & 1664 & 0.660 & 0.9 & 0 & $10^{-10}$ & 0.40 \\
Char ($c$)        & 0.065 & 0.435 & 150   & 1219 & 0.283 & 0.9 & \num{3.3e-3} & $10^{-10}$ & 0.80 \\
Ash ($a$)         & 0.058 & 0.353 & 5.7   & 1244 & 0.315 & 0.9 & \num{6.4e-3} & $10^{-10}$ & 0.985 \\
\bottomrule
\end{tabular}
\end{table*}

\subsection{Governing equations}

The governing equations for condensed-phase mass, species and energy, and for gas-phase species and pressure, together with constitutive relations for the heat, mass and momentum fluxes, follow Fourier's, Fick's and Darcy's laws~\cite{Lautenberger2009,Lautenberger2009_FSJ,lautenberger2014gpyro3d}. The evaporation rate $\dot\omega'''_e$ (positive for evaporation, negative for condensation) appears as a source in the gas mass, gas species, pressure and energy equations.
\begin{itemize}
    \item Condensed-phase (dry solid) mass conservation:
        \begin{equation}
            \pdv{\overline{\rho}_d}{t}=-\dot\omega'''_{fg}
        \end{equation}

    \item Condensed-phase species conservation ($i \ne X$):
        \begin{equation}
            \pdv{(\overline{\rho}_dY_i)}{t}=\dot\omega'''_{fi}-\dot\omega'''_{di}
        \end{equation}

    \item Gas-phase mass conservation:
        \begin{equation}
            \pdv{(\rho_g\overline{\psi})}{t}+\nabla{\cdot}\dot{\bm m}''=\dot\omega'''_{fg} + \dot\omega'''_e
        \end{equation}

    \item Gas-phase species conservation:
        \begin{equation}
            \pdv{(\rho_g\overline{\psi}Y_j)}{t}+\nabla{\cdot}(\dot{\bm m}'' Y_j + \dot{\bm{J}}_j'')= 
            % -\,\nabla{\cdot}\dot{\bm{J}}''_j+
            \Delta\dot\omega'''_j + \delta_{j,\ce{H2O}}\,\dot\omega'''_e
        \end{equation}

    \item Combined gas and condensed phase energy conservation:

        \begin{equation}
        \begin{split}
            \pdv{(\overline{\rho}\ h)}{t} ={}& -\nabla{\cdot}\dot{\bm q}''-\dot{Q}'''_{s,g}+\sum_{k=1}^{3}\dot{Q}'''_{s,k}\\
            &+\sum_{i} \Delta\dot\omega'''_i h_i - \dot\omega'''_e\,\Delta h_v(T)
        \end{split}
            \label{eq:energy}
        \end{equation}    

    \item Gas-phase pressure (Darcy) equation:
        \begin{equation}
            \pdv{}{t}\!\left(\frac{P\,M_a}{\mathcal{R}\,T}\,\overline{\psi}\right)
           = \nabla\!\cdot\!\left(\frac{\overline{K}}{\nu}\,\nabla P\right)
           + \dot{\omega}'''_{fg} + \dot\omega'''_e
       \end{equation}

    \item Bound-water (moisture, $i = X$) conservation (\S~\ref{sec:moisture}):
        \begin{equation}
            \frac{\partial X}{\partial t} = -\nabla\cdot\boldsymbol{j}_X - \frac{\dot{\omega}'''_e}{\rho_v}
        \end{equation}
\end{itemize}
Here $\Delta\dot{\omega}'''_{\beta} = \dot{\omega}'''_{f\beta} - \dot{\omega}'''_{d\beta}$ with $\beta \in \{i,j\}$, and $\delta_{j,\ce{H2O}}$ is 1 for water vapor and 0 otherwise. The condensed- and gas-phase source terms account for species formation ($\dot{\omega}'''_{f\beta}$), destruction ($\dot{\omega}'''_{d\beta}$) and species enthalpies $h_i$, with formation-destruction bookkeeping following Ref.~\cite{Lautenberger2009}. The fluxes are
\begin{gather}
\hspace{-0.25in}
\dot{\bm J}''_j = -\,\overline{\psi}\,\rho_g\,\mathcal{D}\,\nabla Y_j,
\quad
\dot{\bm q}'' = -\,\overline{k}\,\nabla T, \quad
\dot{\bm m}'' = -\,\frac{\overline{K}}{\nu}\,\nabla P,
\nonumber\\
\hspace{-0.25in}
\boldsymbol{j}_X = -\left(\mathcal{D}_{X,r}\frac{\partial X}{\partial r},\; \mathcal{D}_{X,\theta}\,\frac{1}{r}\frac{\partial X}{\partial \theta}\right)
\end{gather}
with $\rho_g = P M_a/(\mathcal{R}\,T)$, where $M_a$ is the molar mass of air.

\subsection{Heterogeneous reaction kinetics}

For each pyrolysis or oxidation reaction $k \in \{1,2,3\}$, a condensed-phase reactant $A_k$ is consumed at a volumetric destruction rate $\dot{\omega}'''_{dA_k}$~\cite{Lautenberger2009},
\begin{equation}
\begin{split}
    \dot{\omega}'''_{dA_k}
    ={}&
    \left(
    \frac{\overline{\rho}_d Y_{A_k}}
         {(\overline{\rho}_d Y_{A_k})_{\Sigma}}
    \right)^{n_k}
    (\overline{\rho}_d Y_{A_k})_{\Sigma}\\
    &\times Z_k
    \exp\!\left(-\frac{E_k}{\mathcal{R} T}\right)
    \mathcal{F}_{\ce{O2},k},
\end{split}
    \label{eq:wdot_dAk}
\end{equation}
where $Z_k$ and $E_k$ are the Arrhenius pre-exponential factor and activation energy (Table~\ref{tab:rxn}), respectively, $n_k$ is the reaction order with respect to $A_k$, and $(\overline{\rho}_d Y_{A_k})_{\Sigma}$ is the normalization term defined in Ref.~\cite{Lautenberger2009}. The oxygen sensitivity factor is
\begin{equation}
\mathcal{F}_{\ce{O2},k}
=
\begin{cases}
1, & n_{\ce{O2},k}=0, \\
\min\!\left[(1+Y_{\ce{O2}})^{n_{\ce{O2},k}} - 1,\;1\right], & n_{\ce{O2},k} \neq 0.
\end{cases}
\end{equation}
Thermal pyrolysis has $n_{\ce{O2},k}=0$ because it is oxygen independent. Oxidative pyrolysis and char oxidation have $n_{\ce{O2},k}>0$, which introduces an explicit dependence on the local pore-gas oxygen mass fraction~\cite{Lautenberger2009}.

The condensed-phase product $B_k$ forms according to
\begin{equation}
\dot{\omega}'''_{fB_k}
=
m_{B,k}\,\dot{\omega}'''_{dA_k},
\qquad
m_{B,k}
=
\frac{\rho_{B_k}}{\rho_{A_k}},
\end{equation}
where $\rho_{A_k}$ and $\rho_{B_k}$ are bulk densities, and the total gas generation rate of reaction $k$ is
\begin{equation}
\dot{\omega}'''_{fg,k}
=
\left(1 - m_{B,k}\right)\dot{\omega}'''_{dA_k}.
\end{equation}
For each gaseous species $j$, formation and destruction from reaction $k$ follow from the heterogeneous gas-yield matrix entries $y_{s,j,k}$:
\begin{align}
\dot{\omega}'''_{fj,k}
&=
\dot{\omega}'''_{fg,k}\,\max\!\left(y_{s,j,k},0\right), \nonumber\\
\dot{\omega}'''_{dj,k}
&=
-\,\dot{\omega}'''_{fg,k}\,\min\!\left(y_{s,j,k},0\right),
\end{align}
so that positive yields contribute to formation and negative yields to consumption, both as non-negative rates. The net sources are $\dot{\omega}'''_{f\beta}=\sum_k \dot{\omega}'''_{f\beta,k}$ and $\dot{\omega}'''_{d\beta}=\sum_k \dot{\omega}'''_{d\beta,k}$, consistent with the formulation in Ref.~\cite{Lautenberger2009}. To keep the residue yields physical, the bulk densities satisfy $\rho_{B_k} < \rho_{A_k}$, so that $0 < m_{B,k} < 1$.

The yields of the heterogeneous and homogeneous reactions are listed in Table~\ref{tab:yields}.

\begin{table}[!t]
\centering
\caption{Gaseous species yields per unit mass of gas generated (heterogeneous reactions $k$, $y_{s,j,k}$) or of fuel consumed (homogeneous reactions $\ell$, $y_{g,j,\ell}$). Only nonzero yields are shown; negative values denote consumption.}
\label{tab:yields}
\small
\setlength{\tabcolsep}{4pt}
\begin{tabular}{@{}lcccccc@{}}
\toprule
 & \multicolumn{4}{c}{$k$} & \multicolumn{2}{c}{$\ell$} \\
\cmidrule(lr){2-5}\cmidrule(lr){6-7}
$j$ & 1 & 2 & 3 & 4 & 1 & 2 \\
\midrule
1 Thermal pyrolysate     & 1 &        &        &   & $-1$   &        \\
2 \ce{N2}                &   &        &        &   &        &        \\
3 \ce{O2}                &   & $-0.1$ & $-2.0$ &   & $-1.6$ & $-1.5$ \\
4 Oxidative pyrolysate   &   & 1.1    &        &   &        & $-1$   \\
5 Char-oxidation prod.   &   &        & 3.0    &   &        &        \\
6 Pyrolysate-oxid. prod. &   &        &        &   & 2.6    & 2.5    \\
7 \ce{H2O}               &   &        &        & 1 &        &        \\
\bottomrule
\end{tabular}
\end{table}

\subsection{Homogeneous gas-phase reactions}

Homogeneous reactions in the pore gas are also considered. Following Ref.~\cite{Lautenberger2009}, each homogeneous reaction $\ell$ involves a fuel species $A_\ell$ and an oxidizer $B_\ell$, and its rate is expressed through the volumetric destruction rate of the fuel,
\begin{equation}
    \dot{\omega}'''_{dA_\ell}
    = [A_\ell]^{p_\ell}\,[B_\ell]^{q_\ell}\,
      T^{b_\ell}\, Z_\ell\, \overline{\psi}\,
      \exp\!\left(-\frac{E_\ell}{\mathcal{R}\,T}\right),
    \label{eq:hom_rate}
\end{equation}
where $[A] = \rho_g Y_A / M_A$ is the molar concentration of gaseous species $A$. The rates of all other species in reaction $\ell$ follow from the homogeneous yield matrix, $\dot{\omega}'''_{g,j,\ell} = \dot{\omega}'''_{dA_\ell}\, y_{g,j,\ell}$~(Table~\ref{tab:yields}). The two reactions included are oxidation of the thermal pyrolysate and oxidation of the oxidative pyrolysate.

\subsection{Heat release terms}

In Eq.~\eqref{eq:energy}, $\dot{Q}'''_{s,k}$ is the volumetric heat release to the condensed phase from reaction $k$ and $\dot{Q}'''_{s,g}$ is the volumetric heat transfer from the condensed phase to the gas phase~\cite{Lautenberger2009}:
\begin{align}
    \dot{Q}'''_{s,k} &= -\dot{\omega}'''_{fB_k}\,\Delta H_{sol,k} - \dot{\omega}'''_{fg,k}\,\Delta H_{vol,k}, \\[4pt]
    \dot{Q}'''_{s,g} &= c_{p,g}\,\dot{\bm m}''\cdot\nabla T - \sum_{\ell=1}^{L}\dot{Q}'''_{g,\ell},
\end{align}
where $\Delta H_{sol,k}$ and $\Delta H_{vol,k}$ are the heats of reaction for forming condensed- and gas-phase species by reaction $k$, and
\begin{equation}
    \dot{Q}'''_{g,\ell} = -\dot{\omega}'''_{dA_\ell}\,\Delta H_\ell
\end{equation}
is the volumetric heat release by homogeneous reaction $\ell$. The last term of Eq.~\eqref{eq:energy} is the latent heat absorbed by evaporation or released by condensation.

Kinetic parameters and species properties follow Ref.~\cite{Lautenberger2009}, except for the bulk densities of virgin wood and char and the emissivities, which are adjusted for white oak. The reaction parameters are listed in Table~\ref{tab:rxn}.

\begin{table*}[!t]
\centering
\caption{Reaction parameters. Heterogeneous parameters, except the bulk densities that set the residue yields $m_{B,k}$, and homogeneous parameters are from Ref.~\cite{Lautenberger2009}. Positive heats of reaction are endothermic.}
\label{tab:rxn}
\small
\begin{tabular}{@{}clcccccc@{}}
\toprule
\multicolumn{8}{@{}l}{(a) Heterogeneous reactions} \\
\midrule
$k$ & From $\to$ To & $\Delta H_{sol}$ (\si{\joule\per\kilogram}) & $\Delta H_{vol}$ (\si{\joule\per\kilogram}) & $Z$ (\si{\per\second}) & $E$ (\si{\kilo\joule\per\mole}) & $n$ & $n_{\ce{O2}}$ \\
\midrule
1 & virgin wood $\to$ char  & 0  & \num{5.33e5}  & \num{3.29e9}  & 135.0 & 4.78 & 0    \\
2 & virgin wood $\to$ char  & 0  & \num{-9.94e5} & \num{6.00e9}  & 124.2 & 4.99 & 1.16 \\
3 & char $\to$ ash          & 0  & \num{-3.77e7} & \num{9.79e13} & 192.4 & 1.86 & 1.04 \\
4 & bound water $\to$ vapor & -- & $\Delta h_v(T)$ & \multicolumn{4}{c}{Evaporation rate computed using Eq.~\eqref{eq:evap}} \\
\midrule
\multicolumn{8}{@{}l}{(b) Homogeneous reactions, $p_\ell = q_\ell = 1$, $b_\ell = 0$} \\
\midrule
$\ell$ & Fuel $A_\ell$ & Oxidizer $B_\ell$ & $\Delta H_\ell$ (\si{\joule\per\kilogram}) & $Z_\ell$ (\si{\kilogram\meter\cubed\per\mole\squared\per\second}) & $E_\ell$ (\si{\kilo\joule\per\mole}) & & \\
\midrule
1 & thermal pyrolysate   & \ce{O2} & \num{-1.88e7} & \num{2.01e9} & 162.7 & & \\
2 & oxidative pyrolysate & \ce{O2} & \num{-1.73e7} & \num{2.09e9} & 161.3 & & \\
\bottomrule
\end{tabular}
\end{table*}

\subsection{Moisture transport and evaporation}
\label{sec:moisture}

\subsubsection{Bound-water transport}

In polar coordinates the moisture equation is given as
\begin{equation}
\begin{split}
    \pdv{X}{t}
    ={}& \frac{1}{r}\pdv{}{r}\!\left(r\,\mathcal{D}_{X,r}\pdv{X}{r}\right)
    + \frac{1}{r^2}\pdv{}{\theta}\!\left(\mathcal{D}_{X,\theta}\pdv{X}{\theta}\right)\\
    &- \frac{\dot\omega'''_e}{\rho_v},
\end{split}
    \label{eq:moisture}
\end{equation}
where $\mathcal{D}_{X,r}$ and $\mathcal{D}_{X,\theta}$ are the effective moisture diffusivities in the radial and tangential directions, related by a user-specified anisotropy ratio, $\mathcal{D}_{X,\theta} = a_X\,\mathcal{D}_{X,r}$. All simulations in this work use $a_X = 1$. The radial diffusivity follows a series model~\cite{gezici2017bound}, in which bound water crosses the cell walls and lumens in series~\cite{siau2012transport},
\begin{align}
    \mathcal{D}_{X,r} &= \frac{\mathcal{D}_b(X,T)}{\xi_w}, \nonumber\\
    \mathcal{D}_b &= \mathcal{D}_{b,0}\exp\!\left(-\frac{E_b}{\mathcal{R}T}\right), \label{eq:siau}\\
    E_b &= E_{b,0} - E_{b,1}\min(X, X_{\mathrm{fsp}}), \nonumber
\end{align}
where $\mathcal{D}_b$ is the bound-water (cell-wall) diffusivity of Siau~\cite{siau2012transport} with $\mathcal{D}_{b,0} = \SI{7e-6}{\meter\squared\per\second}$, $E_{b,0} = \SI{38.5}{\kilo\joule\per\mole}$ and $E_{b,1} = \SI{29}{\kilo\joule\per\mole}$, $\xi_w = 0.16$ is the fraction of the transport path that lies in cell walls~\cite{gezici2017bound}, and $X_{\mathrm{fsp}} = 0.28$ is the fiber saturation point, above which the cell walls are saturated and $\mathcal{D}_b$ no longer increases. Water also moves through the pores as vapor, by diffusion and Darcy flow in the gas-species equations (Section~\ref{sec:evap}). The two paths act in parallel and are linked by the evaporation-condensation exchange.

Bound water does not cross the outer surface directly. It reaches the surroundings only after evaporating. The conductivity and specific heat of water are evaluated as fourth-order polynomials in temperature~(Table~\ref{tab:water}),
\begin{equation}
    k_w = \sum_{m=0}^{4} a_m T^m, \qquad c_{p,w} = \sum_{m=0}^{4} b_m T^m,
\end{equation}
with $T$ in \si{\kelvin}, $k_w$ in \si{\watt\per\meter\per\kelvin}, $c_{p,w}$ in \si{\joule\per\kilogram\per\kelvin}, and the coefficients in Table~\ref{tab:water}.

\begin{table}
\centering
\caption{Polynomial coefficients for the conductivity and specific heat of water~\cite{aspenhysys}.}
\label{tab:water}
\small
\begin{tabular}{@{}crr@{}}
\toprule
$m$ & $a_m$ & $b_m$ \\
\midrule
0 & \num{5.01189}      & \num{3.18591e4} \\
1 & \num{-6.44535e-2}  & \num{-3.260186e2} \\
2 & \num{3.25465e-4}   & \num{1.441786} \\
3 & \num{-6.91901e-7}  & \num{-2.841073e-3} \\
4 & \num{5.33818e-10}  & \num{2.108052e-6} \\
\bottomrule
\end{tabular}
\end{table}

\subsubsection{Equilibrium vapor density}

The water-vapor density in equilibrium with bound water at moisture content $X$ and temperature $T$ is
\begin{equation}
    \rho_{v,s}(X,T) = \phi(X,T)\,\frac{P_{\mathrm{sat}}(T)\,M_w}{\mathcal{R}\,T},
    \label{eq:rhovs}
\end{equation}
where $M_w$ is the molar mass of water, and the saturation vapor pressure is determined using the Clausius-Clapeyron relation~\cite{law2010combustion},
\begin{equation}
    P_{\mathrm{sat}}(T) = P_{\mathrm{atm}}\exp\!\left[-\frac{\Delta h_v(T)\,M_w}{\mathcal{R}}\left(\frac{1}{T}-\frac{1}{T_b}\right)\right],
\end{equation}
with $T_b$ the normal boiling point. The water activity $\phi(X,T)$, the equilibrium relative humidity at moisture content $X$ and temperature $T$, is obtained by numerically inverting the Hailwood-Horrobin sorption isotherm~\cite{hailwood1946absorption} in the form given in the Wood Handbook~\cite{ross2010wood},
\begin{equation}
    X = \frac{18}{W}\left[\frac{H\phi}{1-H\phi} + \frac{H_1H\phi + 2H_1H_2H^2\phi^2}{1 + H_1H\phi + H_1H_2H^2\phi^2}\right],
    \label{eq:hh}
\end{equation}
with
\begin{align}
    W &= 349 + 1.29\,T_c + 0.0135\,T_c^2, \nonumber\\
    H &= 0.805 + 7.36\times10^{-4}\,T_c - 2.73\times10^{-6}\,T_c^2, \nonumber\\
    H_1 &= 6.27 - 9.38\times10^{-3}\,T_c - 3.03\times10^{-4}\,T_c^2, \nonumber\\
    H_2 &= 1.91 + 4.07\times10^{-2}\,T_c - 2.93\times10^{-4}\,T_c^2, \nonumber
\end{align}
where $T_c$ is the temperature in \si{\celsius}. When $X$ exceeds the isotherm value at $\phi = 1$ for the local temperature (0.28 at \SI{37.8}{\celsius}), $\phi = 1$ is used.

\subsubsection{Evaporation rate}
\label{sec:evap}

Evaporation is a finite-rate exchange driven by the difference between the equilibrium vapor density, Eq.~\eqref{eq:rhovs}, and the vapor density of the gas in contact with the solid,
\begin{equation}
    \dot\omega'''_e = \beta\left(\rho_{v,s} - \rho_{v}^{*}\right),
    \label{eq:evap}
\end{equation}
where $\beta$ is a volumetric exchange coefficient (\si{\per\second}). A negative $\dot\omega'''_e$ represents condensation.

\textit{Interior nodes}: The cell walls exchange vapor with the pore gas of the same node, $\rho_v^{*} = \rho_g Y_{\ce{H2O}}$. The pores are represented as the voids of a bed of fibers of diameter $d_f$, which gives the interfacial area per unit volume and the exchange coefficient
\begin{equation}
    A''' = \frac{4(1-\overline{\psi})}{d_f}, \qquad
    h_d = \frac{\mathrm{Sh}_p\,\mathcal{D}_v}{d_f},
\end{equation}
so that $\beta = h_d A'''$.
The pore Sherwood number correlation for packed beds~\cite{leach2000kinetic,wakao1979effect}, with a factor $C_b$ for beds of cylinders is given as,
\begin{equation}
    \mathrm{Sh}_p = C_b\left(2 + 1.1\,\mathrm{Re}_p^{0.6}\,\mathrm{Sc}^{1/3}\right), \qquad
    \mathrm{Re}_p = \frac{\dot m''\,d_f}{\mu_g},
\end{equation}
where $\dot m''$ is the magnitude of the local superficial Darcy mass flux and $\mathrm{Sc} = \mu_g/(\rho_g\mathcal{D}_v)$. The fiber diameter is $d_f = \SI{30}{\micro\meter}$~\cite{sousa2009characterization}, typical of hardwood fibers and $C_b = 0.79$. 
Every quantity in $\beta$ is a property of the wood and the pore gas.
% , so $\beta$ does not depend on the computational grid.

\textit{Surface nodes}: The surface node exchanges directly with the free stream through its outer face, of area $A_s = R\,\Delta\theta\,L$, so that $\beta = h_{d,s}A_s/V$, with
\begin{gather}
    \rho_v^{*} = \rho_{v,\infty} = \phi_\infty\,\frac{P_{\mathrm{sat}}(T_\infty)\,M_w}{\mathcal{R}\,T_\infty}, \nonumber\\
    h_{d,s}(\theta) = \frac{\mathrm{Sh}(\theta)\,\mathcal{D}_v}{D},
    \label{eqn:hd_drying}
\end{gather}
where $\phi_\infty$ is the relative humidity of the free stream, $V$ the volume of the surface node and $D = 2R$ the cylinder diameter. The local Sherwood number follows from the same angular distribution as the local Nusselt number (Section~\ref{sec:bc}), through the heat-mass transfer analogy,
\begin{equation}
    \mathrm{Sh}(\theta) = \mathcal{G}(\theta)\,\mathrm{Re}^{1/2}\,\mathrm{Sc}^{1/3},
    \label{eq:sh_theta}
\end{equation}
with $\mathrm{Re} = \rho_g U D/\mu_g$ and $\mathrm{Sc}$ evaluated at the film temperature $T_\text{f} =(T_\text{g} +  \frac{1}{2\pi}\int T_s(\theta)\dd\theta)/2$.

The binary diffusivity of water vapor in air is~\cite{grigoriev1988thermal,bhattacharjee2024modeling,bhattacharjee2025multiscale}
\begin{equation}
    \mathcal{D}_v = \mathcal{D}_{v,0}\,(2\tilde T)^{1.75},
\end{equation}
where $\mathcal{D}_{v,0} = \SI{3.564e-10}{\meter\squared\per\second\per\kelvin\tothe{1.75}}$, with $\tilde T = T$ within the solid and $\tilde T = (T_s + T_\infty)/2$ at the surface. The free (binary) diffusivity is used in both exchange coefficients: inside the solid the vapor crosses a single pore, and at the surface the Sherwood correlation describes the external boundary layer. Transport of vapor through the pore network is a separate process, described by the species equation with the effective diffusivity $\mathcal{D}_v/\overline{\tau}$ and the porosity $\overline{\psi}$ in the diffusive flux $-\overline{\psi}\rho_g(\mathcal{D}_v/\overline{\tau})\nabla Y_{\ce{H2O}}$. The tortuosity of each condensed species is $\tau_i = 1 + 0.01\,\rho_i$, with $\rho_i$ in \si{\kilogram\per\meter\cubed}~\cite{wassilieff1996sound}.

\subsubsection{Initial conditions}

The solid is initially virgin wood ($\chi_v = 1$) at uniform temperature $T_0$ and moisture content $X_0$, and the pore pressure is atmospheric. The pore gas starts as air in equilibrium with the wet solid,
\begin{equation}
    Y_{\ce{H2O}}(t=0) = \frac{\rho_{v,s}(X_0,T_0)}{\rho_g(T_0)},
\end{equation}
which avoids a spurious burst of evaporation at $t = 0$ as every node fills its pores with vapor.

\subsection{Oxygen-availability limiting of oxidative reactions}

Four reactions consume gas-phase oxygen: oxidative pyrolysis, char oxidation, and the two homogeneous pyrolysate-oxidation reactions. Their rates are evaluated from the state at the start of each time step, so their combined consumption over a step is not intrinsically bounded by the oxygen present in the local pore gas. Left unconstrained, this can drive the local oxygen mass fraction negative and over-predict the exothermic heat release. A local oxygen-budget limiter is therefore applied at every node. The oxygen demand is
\begin{equation}
    \dot{D}_{\ce{O2}}
    =
    \sum_{k=2}^{3}\left|y_{s,\ce{O2},k}\right|\dot{\omega}'''_{fg,k}
    +
    \sum_{\ell=1}^{L}\left|y_{g,\ce{O2},\ell}\right|\dot{\omega}'''_{dA_\ell},
    \label{eq:o2_demand}
\end{equation}
and the admissible supply is the total amount of the oxygen held in the local pore volume,
\begin{equation}
    \dot{S}_{\ce{O2}}
    =
    \frac{\overline{\psi}\,\rho_g\,Y_{\ce{O2}}}{\Delta t},
    \label{eq:o2_supply}
\end{equation}
When demand exceeds supply, all oxygen-consuming rates are multiplied by $\lambda = \dot{S}_{\ce{O2}}/\dot{D}_{\ce{O2}}$, and the species yields and heat release are recomputed from the rescaled rates. The uniform factor allocates the available oxygen proportionally among the competing reactions, and in oxygen-depleted regions it makes the oxidative reactions locally transport-limited rather than kinetics-limited. 

\subsection{Boundary conditions}
\label{sec:bc}

With the outward unit normal vector, $\hat{\bm n}$, at $r = R$, the energy and gas-species boundary conditions balance the net flux into the solid:
\begin{align}
    &\overline{k}\,\nabla T\cdot\hat{\bm n} = h_c(\theta)\left(T_\infty - T_s\right) + \overline{\varepsilon}\left[G(\theta) - \sigma T_s^4\right], \label{eq:bc_T}\\
    &\overline{\psi}\rho_g\mathcal{D}\,\nabla Y_j\cdot\hat{\bm n} = \frac{h_c(\theta)}{c_{p,g}}\left(Y_{j,\infty}-Y_{j,s}\right). \label{eq:bc_Y}
\end{align}
With this sign convention, a positive right-hand side heats the solid, and oxygen diffuses inward when $Y_{\ce{O2},\infty} > Y_{\ce{O2},s}$. Following Ref.~\cite{Lautenberger2009}, Eq.~\eqref{eq:bc_Y} assumes a unit Lewis number and uses $h_c/c_{p,g}$ for all species. The ambient gas has $Y_{\ce{O2},\infty} = 0.24$ and $Y_{\ce{H2O},\infty} = \rho_{v,\infty}/\rho_{g,\infty}$, which is zero for dry air; the remainder is \ce{N2}. The pressure at $r = R$ equals the ambient pressure, $P_\infty$.

\textit{Moisture}: Bound water only leaves the solid after evaporation, so at the surface the moisture flux equals the evaporation flux to the ambient:
\begin{equation}
    -\rho_v\,\mathcal{D}_{X,r}\pdv{X}{r}\bigg|_{r=R} = h_{d,s}(\theta)\left[\rho_{v,s}(X_s,T_s) - \rho_{v,\infty}\right].
    \label{eq:bc_X}
\end{equation}
Equation~\eqref{eq:bc_X} is applied once, as the evaporation sink of the surface cell (Section~\ref{sec:evap}). The vapor produced enters the pore gas of that cell and leaves the solid through Eq.~\eqref{eq:bc_Y}.

\begin{figure}
    \centering
    \includegraphics[width=\linewidth]{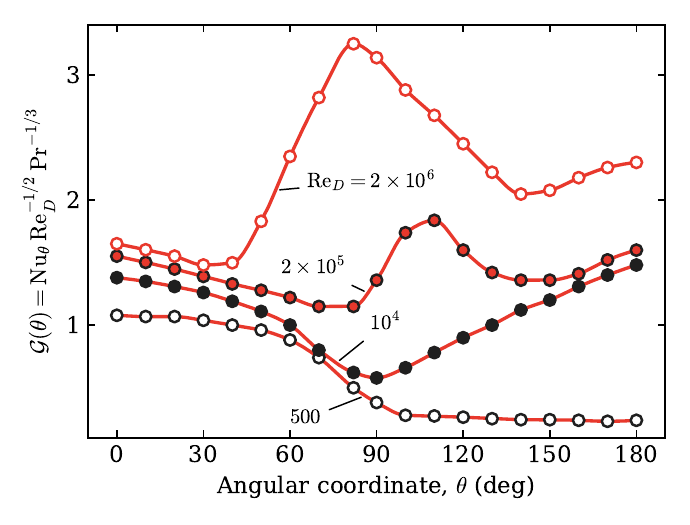}
    \caption{Angular distribution of the local Nusselt number around a cylinder in cross flow, $\mathcal{G}(\theta) = \mathrm{Nu}_\theta\,\mathrm{Re}^{-1/2}\,\mathrm{Pr}^{-1/3}$, for several Reynolds numbers; $\theta = 0$ is the front stagnation point. Adapted from Ref.~\cite{Zukauskas1987}.}
    \label{fig:zukauskas}
\end{figure}

\textit{Convection}: The local heat transfer coefficient follows an angular distribution of the local Nusselt number,
\begin{equation}
    h(\theta) = \mathcal{G}(\theta)\,\mathrm{Re}^{1/2}\,\mathrm{Pr}^{1/3}\,\frac{k_{g,f}}{D},
\end{equation}
where $\mathcal{G}(\theta) = \mathrm{Nu}_\theta\,\mathrm{Re}^{-1/2}\,\mathrm{Pr}^{-1/3}$ is fitted to measured local Nusselt number distributions around a cylinder in cross flow (Fig.~\ref{fig:zukauskas}), $\mathrm{Re}$ and $\mathrm{Pr}$ are evaluated at the mean film temperature for the surface $T_\text{f} =(T_\text{g} +  \frac{1}{2\pi}\int T_s(\theta)\dd\theta)/2$, and $k_{g,f}$ at the local film temperature~\cite{Zukauskas1987}.

Outflow of pyrolysis gases and vapor thickens the boundary layer, which is accounted for by the blowing correction~\cite{Mills2001,Lautenberger2009_FSJ}
\begin{equation}
    h_c(\theta) = \frac{\dot m''_s(\theta)\,c_{p,g}}{\exp\!\left( \dot m''_s(\theta)\,c_{p,g}/h(\theta)\right) - 1}
\end{equation}

so that $h_c(\theta) \to h(\theta)$ as $\dot m''_s(\theta) \to 0$. The correction is applied in all simulations, but during drying the outflow is small and $h_c \approx h$. It becomes significant only once pyrolysis begins in the ignition cases. The surface evaporation coefficient $h_{d,s}$ (Eq.~\eqref{eqn:hd_drying}) is not corrected for blowing.

\textit{Radiation}: The irradiation received by the sample, $G(\theta)$ depends on the configuration and thus is implemented differently for the drying study and the ignition study. 

For the drying study the irradiation to the sample is set to $\sigma T_\infty^4$ and has a negligible contribution. 

In the ignition study, the irradiation is modeled based on the experimental configuration described in Ref.~\cite{Valdivia_thesis} (Fig.~\ref{fig:rad_schematic}). The total irradiation received by each surface element of the wooden dowel accounts for contributions from the radiant panel, the heated surfaces of the flow supply outlet, and the surrounding laboratory:

\begin{figure}
    \centering
    \includegraphics[width=1.0\linewidth]{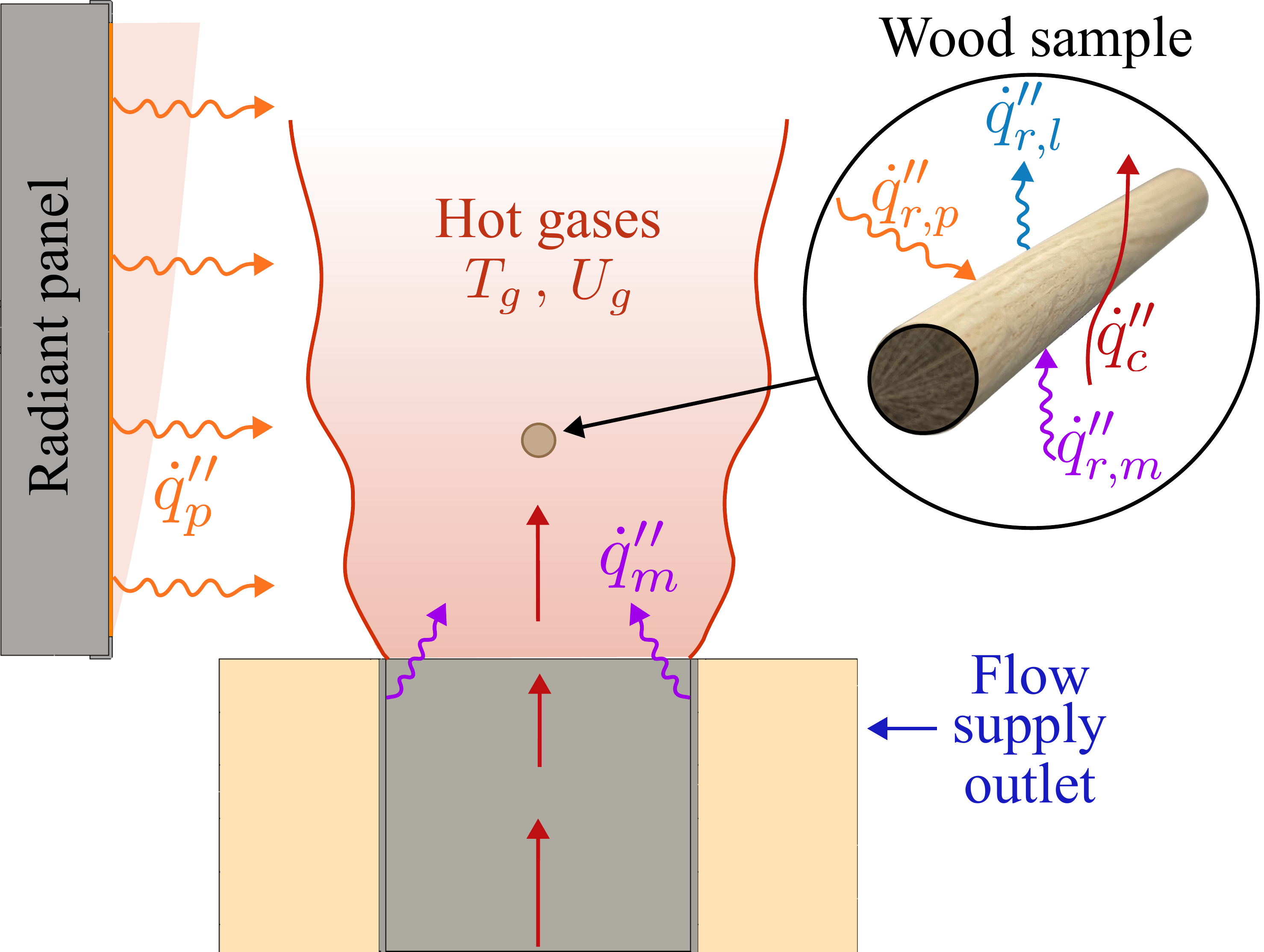}
    \caption{Schematic of radiative and convective heat transfer to a cylindrical wood sample exposed to hot gases and an external radiant panel. The inset illustrates the heat flux contributions at the sample surface. Not to scale. Adapted from Ref.~\cite{Valdivia_thesis}.}
    \label{fig:rad_schematic}
\end{figure}
\begin{equation}
    G(\theta) = {F_{\mathrm{d}S_{\theta}\to p}\,\dot q''_p} + {F_{\mathrm{d}S_{\theta}\to m}\dot q''_m} + F_{\mathrm{d}S_{\theta}\to \mathrm{lab}}\sigma T_{\mathrm{lab}}^4,%\\
\end{equation}
with,
\begin{equation*}
    F_{\mathrm{d}S_{\theta}\to \mathrm{lab} }= 1- F_{\mathrm{d}S_{\theta}\to \mathrm{p}} - F_{\mathrm{d}S_{\theta}\to \mathrm{m}}. 
\end{equation*}
Here, $F_{\mathrm{d}S_{\theta}\to p}$, $F_{\mathrm{d}S_{\theta}\to m}$ and $F_{\mathrm{d}S_{\theta}\to \mathrm{lab}}$ are the view factors from a small surface element of the dowel at angle $\theta$ to the radiant panel, heated surfaces of the flow supply outlet, and laboratory surroundings, respectively~\cite{Valdivia_thesis}. The quantities $\dot q''_p$ and $\dot q''_m$ represent the emitted radiative fluxes from the radiant panel and heated metallic surfaces of the flow supply outlet, respectively, while $T_{\mathrm{lab}}$ denotes the laboratory temperature. Note that $\dot q''_{r,l}=\overline{\varepsilon}\sigma T_s^4$ in Fig.~\ref{fig:rad_schematic} represents the radiative losses. All incident fluxes are angular dependent.

\textit{Center}: At the center ($r = 0$), all fields must be finite and single-valued, which is imposed through Eq.~\eqref{eq:pole_condition}.

%%%%%%%%%%%%%%%%%%%%%%%%%%%%%%%%%%%%%%%%%%%%%%%%%%%%%%%%%%%%%%%%%%%%%
\section{Numerical method}
\label{sec:numerics}

\subsection{Spatial discretization}

The cylinder cross-section is discretized on a uniform two-dimensional polar grid $(r,\theta)$, as illustrated in Fig.~\ref{fig:schematic}. The radial coordinate is divided into $N_r$ nodes between the axis ($r=0$) and the surface ($r=R$), with spacing $\Delta r = R/(N_r-1)$, and the angular coordinate into $N_\theta$ nodes with $\Delta\theta = 2\pi/N_\theta$ and periodic connectivity.

The governing equations are solved with a finite-difference method. For temperature and gas species, diffusive terms are discretized with second-order central differences applied to the expanded form of the polar Laplacian,
\begin{equation}
    \nabla^2 \varphi =
    \frac{\partial^2 \varphi}{\partial r^2}
    + \frac{1}{r}\frac{\partial \varphi}{\partial r}
    + \frac{1}{r^2}\frac{\partial^2 \varphi}{\partial \theta^2},
    \label{eq:polar_laplacian}
\end{equation}
spatially varying transport coefficients are differentiated by product-rule expansion with central differences, and Darcy-driven advection terms use first-order upwinding according to the sign of the local pressure gradient. 

At the coordinate singularity ($r=0$), a single value is computed by constraining the axis node to the angular mean of the first interior ring,
\begin{equation}
    \varphi(0) = \frac{1}{N_\theta}\sum_{j=1}^{N_\theta}\varphi(\Delta r, \theta_j),
    \label{eq:pole_condition}
\end{equation}
which enforces single-valuedness at the axis. At the surface ($r=R$), boundary conditions are imposed through discrete flux balances at the boundary node, equivalent to second-order ghost-node elimination. For moisture, Eq.~\eqref{eq:bc_X} is applied as an evaporation sink in the surface cell. Surface radiative emission is linearized about the start-of-step surface temperature.

Volumetric source terms are evaluated per node using the volume of the annular sector associated with each grid point.

\subsection{Meshing strategy}

All diameters use the same number of radial nodes (fixed $\Delta r/R$), chosen from the mesh convergence study in Section~\ref{sec:mesh}.

\subsection{Time integration and solution sequence}

Energy and moisture are advanced with the Crank-Nicolson scheme ($\vartheta = 0.5$). Pressure and gas species are advanced with the backward Euler scheme ($\vartheta = 1$), which damps their fast transients so that the time step is not restricted by them. The equations are solved sequentially within each step $t^n \to t^{n+1}$:
\begin{enumerate}
    \item The reaction rates, the surface evaporation rate, and the interior equilibrium vapor densities $\rho_{v,s}$ and exchange coefficients $\beta$ are evaluated from the start-of-step state, and the condensed-phase composition is advanced to $\chi_i^{n+1}$. The Darcy mass flux used in $\mathrm{Re}_p$ is taken from the previous step.
    \item The gas-species equations are solved for $Y_j^{n+1}$. The interior evaporation rate, Eq.~\eqref{eq:evap}, is evaluated with the end-of-step vapor density $\rho_g Y_{\ce{H2O}}^{n+1}$, so the exchange is implicit in the vapor equation.
    \item The moisture equation is solved for $X^{n+1}$ with the evaporation rate realized in step 2, so the water lost by the cell walls equals the vapor gained by the pore gas.
    \item The energy equation is solved for $T^{n+1}$.
    \item The pressure equation is solved for $P^{n+1}$.
    \item Properties are updated from $(T^{n+1},\chi_i^{n+1},X^{n+1})$.
\end{enumerate}
Source terms in the energy equation are computed as the $\vartheta$-weighted average of the source fields of the previous and current steps. The resulting sparse linear systems are solved directly. 

\subsection{Time-step selection}
\label{sec:dt}

The time step is selected differently for the drying and ignition cases, which are governed by very different time scales.

\textit{Drying cases}: Because the implicit schemes are unconditionally stable, the time step is chosen for accuracy. After each step, the largest changes in temperature, moisture content and vapor mass fraction are compared with target values,
\begin{equation}
    e = \max\!\left(
    \frac{\lVert \Delta T\rVert_\infty}{\delta_T},\;
    \frac{\lVert \Delta X\rVert_\infty}{\delta_X},\;
    \frac{\lVert \Delta Y_{\ce{H2O}}\rVert_\infty}{\delta_Y}\right),
\end{equation}
with $\delta_T = \SI{0.5}{\kelvin}$, $\delta_X = \num{2e-3}$ and $\delta_Y = 0.02$. If $e > 2$, the step is rejected and repeated with half the time step. Otherwise it is accepted, and the next time step is
\begin{equation}
    \Delta t^{n+1} = \Delta t^{n}\,\min\!\left[1.5,\; \max\!\left(0.3,\; \frac{0.9}{e}\right)\right],
\end{equation}
limited to $\SI{1e-4}{\second} \le \Delta t \le \SI{300}{\second}$. Because the temperature is uniform, the time step is set by the moisture and vapor targets.

\textit{Ignition cases}: The time step is set each step as $\Delta t = \min(\Delta t_T, \Delta t_{\mathrm{rxn}})$, with a diffusive bound scaled on the smallest grid spacing, the arc length $\Delta r\,\Delta\theta$ of the first interior ring,
\begin{equation}
    \Delta t_T = \frac{C_{\Delta t}}{\alpha_{\max}\,\Lambda}, \qquad
    \Lambda = \frac{1}{\Delta r^{2}} + \frac{1}{(\Delta r\,\Delta\theta)^{2}},
\end{equation}
where $\alpha_{\max} = \max(\overline{k}/(\overline{\rho}\,\overline{c}_p))$ and $C_{\Delta t} = 0.01$, and a reactive bound
\begin{gather}
    \Delta t_{\mathrm{rxn}} = 0.2\,\min_{\frac{\partial\dot S}{\partial T} > 0}\left(\frac{\overline{\rho}\,\overline{c}_p}{\partial \dot S/\partial T}\right), \nonumber\\
    \frac{\partial \dot S}{\partial T} \approx \sum_{k=1}^{3} \dot Q'''_{s,k}\,\frac{E_k}{\mathcal{R}T^2} + \sum_{\ell=1}^{L} \dot Q'''_{g,\ell}\,\frac{E_\ell}{\mathcal{R}T^2},
\end{gather}
which estimates the growth rate of the heat release with temperature. Gas-species transport, whose diffusivity is larger than the thermal diffusivity, is sub-cycled with $N_s = \lceil \Delta t/\Delta t_Y\rceil$ sub-steps of length $\Delta t_s = \Delta t/N_s$, where $\Delta t_Y = C_{\Delta t}/(\mathcal{D}_{\max}\Lambda)$.

\section{Drying time of wooden dowels}
\label{sec:drying}
\subsection{Modeling configuration}

Cylinders of \SIlist{5;15;50;150;500}{\milli\meter} diameter are initialized at $X_0 = 0.10$ and $T_0 = T_\infty$ and exposed to air at \SI{37.8}{\celsius} (\SI{310.9}{\kelvin}), 0\% relative humidity and \SI{22.4}{\meter\per\second}, with no radiant heating. Pyrolysis is negligible at this temperature. The water lost is computed from the total water content of the cylinder, bound water plus pore vapor, relative to its initial value. Because the equilibrium moisture content tends to zero at 0\% relative humidity and is approached only asymptotically, drying is characterized by the timelag $t_{63}$ and by $t_{90}$, the times to lose 63.2\% and 90\% of the initial water. With $X_0 = 0.10$, $t_{90}$ is also the time for the mean moisture content to fall to 1\%.

\subsection{Mesh convergence}
\label{sec:mesh}

Mesh convergence is assessed for the \SI{50}{\milli\meter} cylinder ($R = \SI{25}{\milli\meter}$). Because moisture transport during drying is close to axisymmetric, the angular resolution is held at $N_\theta = 15$ and the radial resolution is refined from $N_r = 100$ to 400 (Table~\ref{tab:mesh}). The quantity of interest is the timelag $t_{63}$, the time at which 63.2\% of the initial water has been lost; $t_{50}$ is also reported. Because the equilibrium moisture content is zero at 0\% relative humidity, $t_{63}$ coincides with the standard definition of the fuel timelag~\cite{Anderson1982}.

The discretization uncertainty is estimated with the grid convergence index (GCI) of Celik et al.~\cite{celik2008procedure}. Each grid is characterized by its radial spacing $h = \Delta r = R/(N_r-1)$. Three grids with a near-constant refinement ratio are used: $N_r = 400$, 200 and 100, denoted by $k = 1$ (fine), 2 (medium) and 3 (coarse), so that $h_1 < h_2 < h_3$ and $r_{21} = h_2/h_1 = 2.01$, $r_{32} = h_3/h_2 = 2.01$. With $\varphi_k$ the value of $t_{63}$ on grid $k$, $\varepsilon_{21} = \varphi_2 - \varphi_1$ and $\varepsilon_{32} = \varphi_3 - \varphi_2$, the apparent order of accuracy $p$ is obtained iteratively from
\begin{equation}
    p = \frac{\left|\ln\left|\varepsilon_{32}/\varepsilon_{21}\right| + \ln\!\left(\dfrac{r_{21}^p - s}{r_{32}^p - s}\right)\right|}{\ln r_{21}},
    \qquad s = \operatorname{sign}\!\left(\frac{\varepsilon_{32}}{\varepsilon_{21}}\right),
\end{equation}
and the extrapolated value and the fine-grid GCI are
\begin{gather}
    \varphi_{\mathrm{ext}} = \frac{r_{21}^p\varphi_1 - \varphi_2}{r_{21}^p - 1}, \nonumber\\
    \mathrm{GCI}_{21} = \frac{1.25\,\left|(\varphi_1-\varphi_2)/\varphi_1\right|}{r_{21}^p - 1}.
\end{gather}

\begin{table}[p]
\centering
\caption{Radial grids for the mesh convergence study ($R = \SI{25}{\milli\meter}$, $N_\theta = 15$) and the times to lose 50\% and 63.2\% of the initial water. $\epsilon$ is the deviation of $t_{63}$ from the extrapolated value $\varphi_{\mathrm{ext}} = \SI{137.7}{\hour}$.}
\label{tab:mesh}
\small
\begin{tabular}{@{}ccccc@{}}
\toprule
$N_r$ & $\Delta r$ (\si{\milli\meter}) & $t_{50}$ (\si{\hour}) & $t_{63}$ (\si{\hour}) & $\epsilon$ (\%) \\
\midrule
100 & 0.253 & 84.9 & 151.5 & 10.0 \\
150 & 0.168 & 82.2 & 146.6 & 6.5 \\
200 & 0.126 & 81.0 & 144.3 & 4.8 \\
250 & 0.100 & 80.2 & 142.9 & 3.8 \\
300 & 0.084 & 79.7 & 142.0 & 3.1 \\
350 & 0.072 & 79.3 & 141.4 & 2.7 \\
400 & 0.063 & 79.1 & 140.9 & 2.3 \\
\bottomrule
\end{tabular}
\end{table}

\begin{figure}
    \centering
    \includegraphics[width=\linewidth]{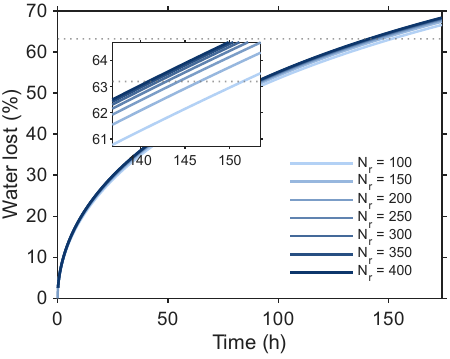}\\[-2pt]
    {\small (a)}\\[4pt]
    \includegraphics[width=\linewidth]{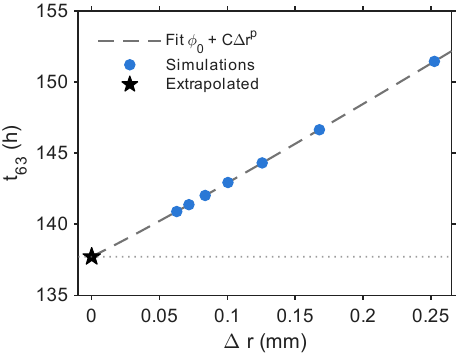}\\[-2pt]
    {\small (b)}
    \caption{Mesh convergence for the \SI{50}{\milli\meter} cylinder ($N_\theta = 15$). (a) Fraction of the initial water lost on the seven grids of Table~\ref{tab:mesh}; the inset shows the 63.2\% crossing. (b) Timelag $t_{63}$ versus radial grid spacing, with a least-squares fit through all seven grids (dashed) and the Richardson-extrapolated value (star).}
    \label{fig:mesh}
\end{figure}

The drying curves converge monotonically as the grid is refined, with coarser grids drying more slowly (Fig.~\ref{fig:mesh}a). The difference between grids is largest near the 63.2\% crossing, where $t_{63}$ falls from \SI{151.5}{\hour} on the coarsest grid to \SI{140.9}{\hour} on the finest (Table~\ref{tab:mesh}, Fig.~\ref{fig:mesh}b). The GCI analysis gives an apparent order of $p = 1.05$, an extrapolated timelag of \SI{137.7}{\hour} and a fine-grid uncertainty of $\mathrm{GCI}_{21} = 2.8\%$, with the grids in the asymptotic range. A least-squares fit through all seven grids gives the same order and extrapolated value. The grid with $N_r = 200$ ($\Delta r/R = 0.005$) predicts $t_{63}$ within 5\% of the extrapolated value and is used for all diameters in Section~\ref{sec:drying_results}.
 
\subsection{Results}
\label{sec:drying_results}

\begin{figure}
    \centering
    \includegraphics[width=\linewidth]{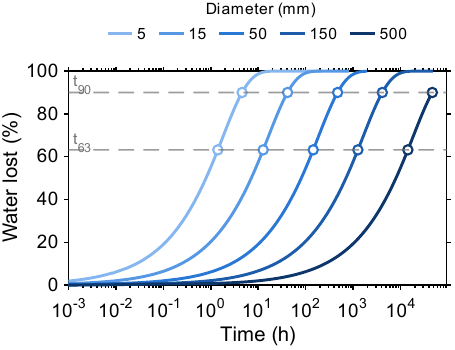}
    \caption{Fraction of the initial water lost by the 5, 15, 50, 150 and \SI{500}{\milli\meter} cylinders. Dashed lines mark 63.2\% ($t_{63}$) and 90\% ($t_{90}$).} 
    \label{fig:drying}
\end{figure}

\begin{table}[p]
\centering
\caption{Predicted drying times and the timelag class of each diameter~\cite{Anderson1982,jolly2024modernizing}.}
\label{tab:drying}
\small
\begin{tabular}{@{}cccc@{}}
\toprule
$D$ (\si{\milli\meter}) & Class & $t_{63}$ (\si{\hour}) & $t_{90}$ (\si{\hour}) \\
\midrule
5   & 1-h    & 1.41              & 4.57 \\
15  & 10-h   & 12.72              & 41.52 \\
50  & 100-h  & 144.3               & 471.6 \\
150 & 1000-h & 1262               & 4129 \\    
500 & 1000-h(+) & \num{1.4e4}    & \num{4.6e4} \\
\bottomrule
\end{tabular}
\end{table}

Figure~\ref{fig:drying} and Table~\ref{tab:drying} compare the drying of the five cylinders. The \SI{5}{\milli\meter} cylinder loses 63.2\% of its water in \SI{1.4}{\hour} and 90\% in \SI{4.6}{\hour}, the \SI{15}{\milli\meter} cylinder in \SI{12.7}{\hour} and \SI{41.5}{\hour}, the \SI{50}{\milli\meter} cylinder in \SI{144}{\hour} and \SI{472}{\hour}, and the \SI{150}{\milli\meter} cylinder in about \SI{1260}{\hour} and \SI{4130}{\hour}. The drying time grows with the square of the diameter: relative to the \SI{50}{\milli\meter} cylinder, the times are scaled by 0.01 for \SI{5}{\milli\meter}, about 8.7 for \SI{150}{\milli\meter} and about 99 for \SI{500}{\milli\meter}, within 5\% of $(D/D_{50})^2$ in every case. This scaling holds even though the Reynolds number changes by two orders of magnitude, which confirms that drying is controlled by moisture diffusion inside the wood rather than by the external flow. Matching $t_{63}$ to the analytical solution for diffusion in a cylinder gives an effective diffusivity of about \SI{1.34e-10}{\meter\squared\per\second}. The ratio $t_{90}/t_{63} \approx 3.26$ is larger than the value of 3.0 obtained from the analytical solution for a cylinder with constant diffusivity~\cite{crank1979mathematics}, because the effective moisture diffusivity decreases as the wood dries.

\begin{figure}
    \centering
    \includegraphics[width=\linewidth]{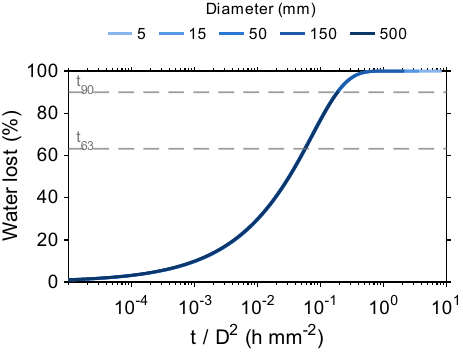}
    \caption{Fraction of the initial water lost against time normalized by the square of the diameter, $t/D^2$, for the 5, 15, 50, 150 and \SI{500}{\milli\meter} cylinders.}
    \label{fig:drying_collapse}
\end{figure}

When plotted against time normalized by the square of the diameter, the fraction of water lost nearly collapses onto a single curve for all five cylinders (Fig.~\ref{fig:drying_collapse}). The collapsed curve reaches 63.2\% at $t/D^2 \approx \SI{0.057}{\hour\per\milli\meter\squared}$, so timelags of 1, 10, 100 and \SI{1000}{\hour} correspond to diameters of about 4, 13, 42 and \SI{132}{\milli\meter}, each within the diameter range of the corresponding timelag class (Section~\ref{sec:intro}). The \SI{150}{\milli\meter} cylinder, a 1000-h fuel, has a timelag of \SI{1260}{\hour}, consistent with its class. The model therefore reproduces the timelag classification of dead fuels from the internal moisture transport properties of the wood alone. The \SI{500}{\milli\meter} cylinder lies well beyond the 1000-h class, and its timelag of about \SI{1.4e4}{\hour} (about 600 days) shows that large logs retain most of their moisture even through a prolonged drought.

\subsection{Effect of air velocity}
\label{sec:drying_velocity}

\begin{figure}
    \centering
    \includegraphics[width=\linewidth]{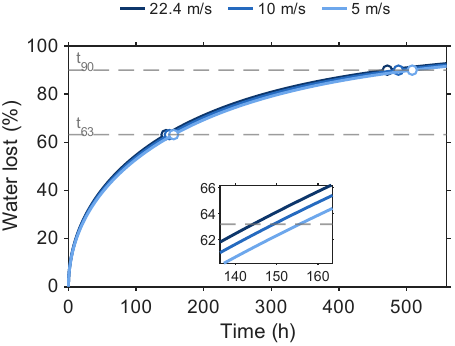}
    \caption{Fraction of the initial water lost by the \SI{50}{\milli\meter} cylinder at three air velocities. Circles mark $t_{63}$ and $t_{90}$; the inset shows the 63.2\% crossing.}
    \label{fig:velocity}
\end{figure}

To see how much the airflow matters, the \SI{50}{\milli\meter} case is repeated at 10 and \SI{5}{\meter\per\second}, with all other conditions unchanged (Table~\ref{tab:velocity}, Fig.~\ref{fig:velocity}).

\begin{table}[p]
\centering
\caption{Drying times of the \SI{50}{\milli\meter} cylinder at three air velocities. $\Delta$ is the increase relative to \SI{22.4}{\meter\per\second}.}
\label{tab:velocity}
\small
\begin{tabular}{@{}ccccc@{}}
\toprule
$U$ (\si{\meter\per\second}) & $Re$ & $t_{63}$ (\si{\hour}) & $t_{90}$ (\si{\hour}) & $\Delta$ (\%) \\
\midrule
22.4 & \num{6.7e4} & 144.3 & 471.6 & -- \\
10   & \num{3.0e4} & 149.3 & 488.0 & 3.5 \\
5    & \num{1.5e4} & 155.5 & 508.4 & 7.8 \\
\bottomrule
\end{tabular}
\end{table}

Lowering the velocity from 22.4 to \SI{5}{\meter\per\second} halves the external mass transfer coefficient, yet $t_{63}$ and $t_{90}$ increase by only 7.8\% relative to the \SI{22.4}{\meter\per\second} case. The drying curves keep the same shape and are only shifted slightly in time. Wind speed therefore has only a small effect on how fast large dead fuels dry, their size and internal transport properties matter much more.

%%%%%%%%%%%%%%%%%%%%%%%%%%%%%%%%%%%%%%%%%%%%%%%%%%%%%%%%%%%%%%%%%%%%%
%%%%%%%%%%%%%%%%%%%%%%%%%%%%%%%%%%%%%%%%%%%%%%%%%%%%%%%%%%%%%%%%%%%%%
\section{Glowing ignition of white oak dowels}
\label{sec:ignition}
 
The model is validated against glowing and flaming ignition times measured for \SI{6.35}{\milli\meter} white oak dowels~\cite{Valdivia_thesis} under two heating modes: convection alone and coupled convection and radiation. Predicted char and ash fronts are compared for coupled heating, and the validated model is then used to examine how the initial moisture content delays glowing ignition under coupled heating.
 
\subsection{Experimental configuration}
\label{sec:exp}
 
The experiments were performed in the Convective-Radiative Ignition Facility~\cite{Valdivia_thesis}, which exposes a horizontal dowel to independently controlled hot-air cross flow and external radiation. The cross flow is supplied by the Convective Ignition Facility~\cite{Valdivia2025}, in which an electric process heater and an insulated plenum with perforated plates and a tube-array flow straightener deliver a uniform air stream at \SIrange{310}{430}{\celsius} and \SIrange{2.1}{2.4}{\meter\per\second}. Radiation is provided by a propane-fired infrared panel with an emissive flux of \SI{86.9}{\kilo\watt\per\meter\squared}; the incident flux, up to about \SI{42}{\kilo\watt\per\meter\squared}, is set by the distance between the panel and the dowel. The heated plenum surfaces contribute up to about 10\% of the panel flux. The dowels were oven-dried at \SI{80}{\celsius} for at least \SI{48}{\hour} and each condition was repeated three times.
 
In the model, the measured gas temperature and velocity define the convective boundary condition, and the panel and plenum emissive fluxes, together with view factors from the dowel surface to each source computed by contour integration (panel) and Monte Carlo ray tracing (plenum and flow straightener)~\cite{Valdivia_thesis}, define the irradiation $G(\theta)$ in Eq.~\eqref{eq:bc_T}. The dowels are taken as dry ($X_0 = 0$). Three conditions are simulated (Table~\ref{tab:ign_cases}): two with combined radiation and convection at different panel distances, and one with convection only. Glowing ignition in the model is defined as the time at which the maximum temperature in the dowel first reaches \SI{1048}{\kelvin}, following Ref.~\cite{Valdivia_thesis}.
 
\begin{table}[p]
\centering
\caption{Validation conditions for the \SI{6.35}{\milli\meter} dowels. $d_p$ is the distance between the radiant panel and the dowel.}
\label{tab:ign_cases}
\small
\begin{tabular}{@{}lccc@{}}
\toprule
Case & $d_p$ (\si{\milli\meter}) & $T_g$ (\si{\celsius}) & $U_g$ (\si{\meter\per\second}) \\
\midrule
1 & 181 & 410 & 2.40 \\
2 & 380 & 428 & 2.44 \\
3 & N/A (panel off) & 428 & 2.44 \\
\bottomrule
\end{tabular}
\end{table}
 
\subsection{Comparison with measured ignition times}
\label{sec:ign_results}
 
\begin{figure}
    \centering
    \subfloat[]{\includegraphics[width=\linewidth]{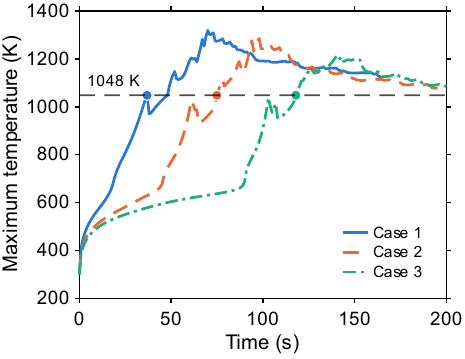}}\\
    \subfloat[]{\includegraphics[width=\linewidth]{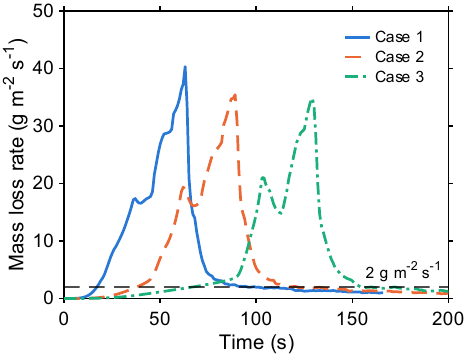}}
    \caption{Predicted (a) maximum temperature and (b) mass loss rate for the cases of Table~\ref{tab:ign_cases}. The dashed lines mark the glowing ignition temperature (\SI{1048}{\kelvin}) and the critical mass flux for flaming ignition (\SI{2}{\gram\per\meter\squared\per\second}); markers show predicted glowing ignition.}
    \label{fig:ign_TM}
\end{figure}

Figure~\ref{fig:ign_TM} shows the predicted maximum temperature and mass loss rate for the three cases. The closer panel in Case~1 heats the dowel fastest, and the maximum temperature crosses \SI{1048}{\kelvin} at \SI{37}{\second}, against \SI{75}{\second} for Case~2 and \SI{118}{\second} for convection alone (Case~3). In every case the temperature continues to rise after glowing ignition, to peaks of \SIrange{1210}{1318}{\kelvin}, driven by the exothermic char oxidation. The mass loss rate rises with temperature, peaks at \SIrange{35}{40}{\gram\per\meter\squared\per\second} and then decays as the virgin wood and char are consumed.
 
\begin{figure}
    \centering
    \subfloat[]{\includegraphics[width=\linewidth]{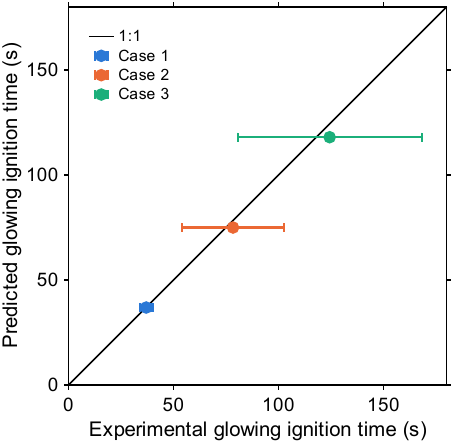}}\\
    \subfloat[]{\includegraphics[width=\linewidth]{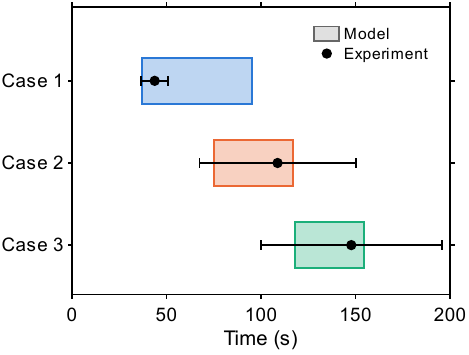}}
    \caption{(a) Predicted against measured glowing ignition time; horizontal bars show one standard deviation of the measurements. (b) Predicted flaming ignition window (shaded) and measured flaming ignition time (mean $\pm$ one standard deviation).}
    \label{fig:ign_val}
\end{figure}

The predicted and measured glowing ignition times are compared in Fig.~\ref{fig:ign_val}a and Table~\ref{tab:ign_results}. The predictions fall within one standard deviation of the measured mean in all three cases. For Case~1 the predicted and measured times are both \SI{37.0}{\second}; for Cases~2 and~3 the model predicts ignition 4\% and 5\% earlier than the measured mean.
 
The model does not resolve the gas phase and therefore cannot predict flaming ignition directly. In the experiments, glowing ignition always preceded flaming ignition~\cite{Valdivia_thesis}, so flaming can only occur after the solid has started glowing and while it still releases enough volatiles to sustain a flame. A model ignition window is therefore defined from the predicted glowing ignition time to the time at which the mass loss rate, after its peak, falls below a critical mass flux of \SI{2}{\gram\per\meter\squared\per\second}~\cite{McAllister2013}. The probability that the measured flaming ignition time lies inside this window is estimated by treating the measurements as normally distributed with the measured mean and standard deviation.
 
\begin{table*}[p]
\centering
\caption{Measured and predicted ignition times for the \SI{6.35}{\milli\meter} dowels. Measured values are mean $\pm$ one standard deviation of three repeats. $P$ is the probability that the measured flaming ignition time lies within the predicted ignition window.}
\label{tab:ign_results}
\small
\begin{tabular}{@{}lccccc@{}}
\toprule
 & \multicolumn{2}{c}{Glowing ignition (\si{\second})} & \multicolumn{2}{c}{Flaming ignition (\si{\second})} & \\
\cmidrule(lr){2-3}\cmidrule(lr){4-5}
Case & Measured & Predicted & Measured & Predicted window & $P$ (\%) \\
\midrule
1 & $37.0 \pm 3.0$   & 37.0 & $43.7 \pm 7.0$   & 37.0-95.4  & 83 \\
2 & $78.3 \pm 24.1$  & 75.0 & $108.7 \pm 41.3$ & 75.0-117.0 & 37 \\
3 & $124.3 \pm 43.8$ & 118.1 & $147.7 \pm 47.9$ & 118.1-154.4 & 29 \\
\bottomrule
\end{tabular}
\end{table*}
 
The measured mean flaming ignition times fall inside the predicted windows for all three cases (Fig.~\ref{fig:ign_val}b). For Case~1, 83\% of the measured distribution lies within the window. The lower probabilities for Cases~2 and~3 (37\% and 29\%) mainly reflect the wide experimental scatter, which extends well past the end of the window; the measured mean for Case~3 lies near the end of its window. Together, the glowing ignition times and ignition windows show that the model captures the effect of the heating mode on ignition: stronger radiation shortens both the time to glowing and the time to flaming, and convection alone produces the slowest ignition.
 
\subsubsection{Decomposition fronts}
\label{sec:ign_fronts}

\begin{figure}
    \centering
    \includegraphics[width=\linewidth]{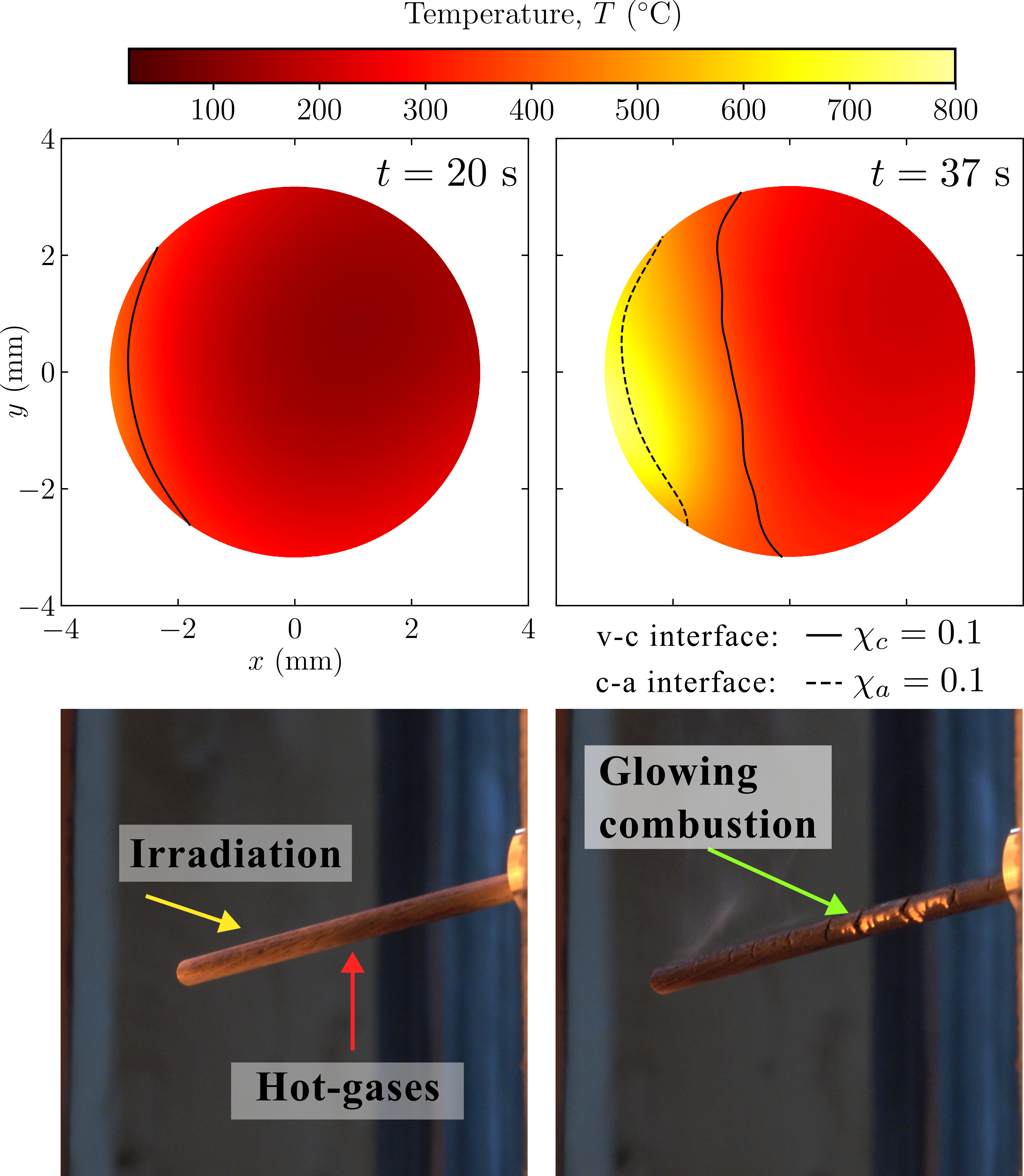}
    \caption{Predicted temperature and decomposition fronts in the \SI{6.35}{\milli\meter} dowel for Case~1 at the first instance of char formation and the mean measured glowing ignition time, with corresponding photographs from the experiments~\cite{Valdivia_thesis}. Solid lines: char front ($\chi_c = 0.1$); dashed lines: ash front ($\chi_a = 0.1$).}
    \label{fig:ign_fronts}
\end{figure}

Figure~\ref{fig:ign_fronts} shows the predicted temperature field and the decomposition fronts for Case~1 (radiation and convection) at the first instance of char formation and predicted glowing ignition. The char and ash fronts are shown by the contours $\chi_c = \chi_a = 0.1$ of the char and ash volume fractions, and are compared with photographs of the dowels from the experiments~\cite{Valdivia_thesis}.

In this case the model predicts the sequence observed in the experiments starting with a char layer formation first at the heated surface, and by the predicted glowing ignition time an ash region has formed on the irradiated side. Under coupled heating (Case~1), the radiation concentrates the heating on the side facing the panel, producing a steep surface temperature gradient and a char front that advances fastest on that side, followed by glowing ignition converting char to ash. 

The predicted glowing ignition time matches the measured mean for Case~1 and is 4-5\% shorter for Cases~2 and~3, well within one standard deviation of the measurements. The remaining differences, and the uncertainty in the ignition windows, may have several causes. First, the kinetic parameters and most thermophysical properties were taken from a model calibrated for white pine~\cite{Lautenberger2009}, whereas the experiments used white oak with only the bulk densities and emissivities were adjusted. Oak differs from pine in density, extractive and lignin content, and char reactivity, all of which affect the onset of char oxidation. Second, the model neglects shrinkage, cracking and ash removal, and treats the gas phase outside the solid only through correlations, so the surface heat and oxygen supply during glowing may be misrepresented. Third, the generalized multi-step mechanism lumps the decomposition chemistry into a few global reactions, which cannot capture every pathway of the real material.
 
\subsection{Effect of initial moisture content}
\label{sec:ign_moisture}

The validated model is used to examine how moisture delays glowing ignition. Case~1 (Table~\ref{tab:ign_cases}) is repeated with initial moisture contents of $X_0 = 0$, 0.05, 0.10, 0.15 and 0.20 (dry basis), all below the fiber saturation point, with all other conditions unchanged. Because the dowels in the experiments were oven-dried, these results are model predictions only.

\begin{figure}
    \centering
    \includegraphics[width=\linewidth]{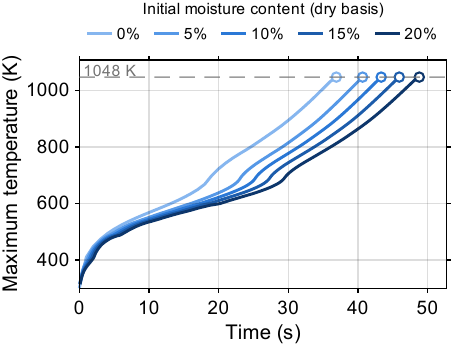}
    \caption{Predicted maximum temperature up to glowing ignition for Case~1 with initial moisture contents $X_0 = 0$-0.20. Each curve ends where it reaches the glowing ignition temperature of \SI{1048}{\kelvin} ($\approx$ \SI{775}{\celsius}) denoted with a dashed line.}
    \label{fig:ign_moisture}
\end{figure}

Figure~\ref{fig:ign_moisture} shows the maximum temperature of the dowel up to glowing ignition. Moisture slows the early temperature rise, because part of the absorbed heat is consumed by evaporation. After \SI{20}{\second}, the dowel with $X_0 = 0.20$ is about \SI{120}{\kelvin} cooler than the dry dowel. Glowing ignition is delayed from \SI{37.0}{\second} for the dry dowel to 40.7, 43.3, 45.9 and \SI{48.8}{\second} for $X_0 = 0.05$, 0.10, 0.15 and 0.20, respectively. The delay grows nearly linearly with moisture content, by about \SI{0.6}{\second} per percentage point, and reaches \SI{11.8}{\second} (32\%) at $X_0 = 0.20$. Even under strong radiative-convective heating, the moisture typical of conditioned dead fuels therefore produces a measurable delay in glowing ignition.

\section{Discussion}
\label{sec:discussion}

\subsection{Model limitations}

The model resolves moisture transport and phase change but simplifies several processes. Liquid water flow above the fiber saturation point is not included, so the formulation applies to dead fuels below about 28\% moisture content. The bound-water diffusivity uses Siau's correlation with a cell-wall path fraction measured for pine sapwood~\cite{gezici2017bound}, and vapor diffusion through the pores acts as a parallel path. The drying times, therefore, depend on both the bound-water diffusivity and the pore properties (porosity and tortuosity). Drying times for a particular species therefore require its own transport data. Axial transport, which is much faster along the grain, is excluded by the two-dimensional geometry, so the results apply to long cylinders whose ends contribute little to the water loss.

The gas phase outside the solid is not resolved. Surface heat and mass transfer come from correlations for a cylinder in cross flow, and flaming ignition is inferred from the mass loss rate rather than predicted. The solid does not shrink, crack or lose ash, which affects the late stages of glowing combustion. Finally, the drying cases use a single, steady weather condition and diurnal cycles in temperature, humidity, radiation and rain are not considered.

\section{Conclusions}
\label{sec:conclusions}

A two-dimensional polar ($r$-$\theta$) model of a moist wood cylinder was developed that couples bound-water  and pore-vapor transport through evaporation-condensation exchange with oxidative pyrolysis and char oxidation under angularly resolved convective and radiative heating. The main findings are:
\begin{enumerate}
    \item Under hot, dry and windy conditions, the drying time of a wood cylinder grows with the square of its diameter, and the curves of water lost against $t/D^2$ collapse. The predicted timelags of 1, 10, 100 and \SI{1000}{\hour} correspond to diameters of about 4, 13, 42 and \SI{132}{\milli\meter}, within the diameter ranges of the standard timelag classes.
    \item Drying is controlled by diffusion of moisture inside the wood, as bound water through the cell walls and as vapor through the pores. Reducing the air velocity from 22.4 to \SI{5}{\meter\per\second} lengthens drying by only 8\%.
    \item For \SI{6.35}{\milli\meter} white oak dowels, the predicted glowing ignition times lie within one standard deviation of the measurements under both coupled convective-radiative heating and convection alone, and the measured mean flaming ignition times fall within the predicted ignition windows.
    \item Under coupled heating, radiation concentrates the heating on the side facing the panel and produces one-sided char and ash fronts, whereas, convection alone gives the slowest ignition.
    \item An initial moisture content of 20\% delays glowing ignition under strong coupled heating by about 30\%, approximately linearly with moisture content.
\end{enumerate}

%%%%%%%%%%%%%%%%%%%%%%%%%%%%%%%%%%%%%%%%%%%%%%%%%%%%%%%%%%%%%%%%%%%%%

\printcredits

\bibliographystyle{elsarticle-num}

\bibliography{references}

\end{document}